\documentclass[12pt,a4paper]{iopart}
\pdfoutput=1
\usepackage{iopams,setstack,mathrsfs,amsthm,cite,enumerate,float,graphicx}
\usepackage[utf8]{inputenc}
\usepackage[T1]{fontenc}
\usepackage{lmodern}
\usepackage[colorlinks,linkcolor=blue,citecolor=blue,urlcolor=blue]{hyperref}
\newcommand{\al}{\alpha}
\newcommand{\be}{\beta}
\newcommand{\de}{\delta}
\newcommand{\ep}{\epsilon}
\newcommand{\vep}{\varepsilon}

\newcommand{\si}{\sigma}

\newcommand{\ze}{\zeta}
\newcommand{\De}{\Delta}

\newcommand{\La}{\Lambda}

\newcommand{\bJ}{\mathbf{J}}
\newcommand{\bk}{\mathbf{k}}

\newcommand{\bp}{\mathbf{p}}

\newcommand{\bt}{\mathbf{t}} 

\newcommand{\bx}{\mathbf{x}}

\newcommand{\bsi}{{\boldsymbol{\si}}}

\newcommand{\CC}{{\mathbb C}}
\newcommand{\NN}{{\mathbb N}}
\newcommand{\RR}{{\mathbb R}}

\newcommand{\cE}{{\mathcal E}}

\newcommand{\cH}{{\mathcal H}}

\newcommand{\cN}{{\mathcal N}}
\newcommand{\cP}{{\mathcal P}}

\newcommand{\cS}{{\mathcal S}}

\newcommand\BJ{\,\overline{\!J\,}}

\newcommand{\HS}{H_{\mathrm{sp}}}
\newcommand{\Hsp}{\hat H_0}
\newcommand\Hsc{H_{\mathrm{sc}}}
\newcommand{\Egs}{E_{\mathrm{GS}}}
\newcommand\Zsc{Z_{\mathrm{sc}}}
\newcommand{\ZS}{Z_{\mathrm{sp}}}

\newcommand{\pd}{\partial}

\newcommand{\ket}[1]{|#1\rangle}

\newcommand{\mss}{\kern 1pt}
\renewcommand{\leq}{\leqslant}

\renewcommand{\le}{\leqslant}

\newcommand{\tends}[1]{\bbuildrel{\hbox to 2em{\rightarrowfill}}_{#1}^{}}

\let\iu\rmi
\let\diff\rmd
\newcommand{\su}{\mathrm{su}}

\newcommand{\qbinom}[3]{{#1\atopwithdelims[]#2}_{\raise 3pt\hbox{$\scriptstyle #3$}}}
\newcommand{\en}{\enspace}

\newcommand{\Int}[1]{\,\mathop{\!#1}\limits^{\lower1ex\hbox{$\scriptstyle\circ$}}{}}

\theoremstyle{remark}

\let\tfrac\case
\let\eqref\eref

\newcommand{\binom}[2]{{#1\choose #2}}
\newcommand{\mathclap}[1]{\hbox to0pt{\hss$\scriptstyle #1$\hss}}

\eqnobysec
\newcommand{\bN}{\mathbf N}

\newcommand{\brv}{\mathbf r}
\newcommand{\bsv}{\mathbf s}

\newcommand{\hatrr}{\hat{r}}
\newcommand{\hatN}{\hat{N}}
\newcommand{\hatbrv}{\hat{\brv}}
\newcommand{\hatbN}{\hat{\bN}}

\newcommand{\Emax}{E_{\mathrm{max}}}
\def\clap#1{\hbox to 0pt{\hss#1\hss}}

\def\mathclap{\mathpalette\mathclapinternal}

\def\mathclapinternal#1#2{%
  \clap{$\mathsurround=0pt#1{#2}$}}

\begin{document}

\title{Generalized Lipkin--Meshkov--Glick models of Haldane--Shastry type}

\author{José A.~Carrasco, Federico Finkel, Artemio González-López}

\address{Departamento de Física Teórica II, Universidad Complutense de Madrid, 28040 Madrid,
  Spain}

\eads{\mailto{joseacar@ucm.es}, \mailto{ffinkel@ucm.es}, \mailto{artemio@ucm.es}}

\date{\today}

\begin{abstract}
  We introduce a class of generalized Lipkin--Meshkov--Glick (gLMG) models with $\su(m)$
  interactions of Haldane--Shastry type. We have computed the partition function of these models
  in closed form by exactly evaluating the partition function of the restriction of a spin chain
  Hamiltonian of Haldane--Shastry type to subspaces with well-defined magnon numbers. As a
  byproduct of our analysis, we have obtained strong numerical evidence of the Gaussian character
  of the level density of the latter restricted Hamiltonians, and studied the distribution of the
  spacings of consecutive unfolded levels. We have also discussed the thermodynamic behavior of a
  large family of $\su(2)$ and $\su(3)$ gLMG models, showing that it is qualitatively similar to
  that of a two-level system.
\end{abstract}

{\it Keywords\/}: integrable spin chains and vertex models, solvable lattice models


\maketitle

\section{Introduction}

One of the first, and still one of the few, quantum mechanical many-body models that has been
solved in the literature is the Lipkin--Meshkov--Glick (LMG) model~\cite{LMG65,MGL65,GLM65}, which
describes a system of $N$ fermions with two $N$-fold degenerate one-particle levels. The original
motivation for introducing this model was testing the validity of different approximation schemes
from solid state physics or field theory in the context of nuclear physics. Over the years, the
LMG model has appeared in connection with a wide range of problems of physical interest, including
shape transitions in nuclei \cite{RS80}, trapped ion and optical cavity
experiments~\cite{UF03,CLJ09}, two-modes Bose--Einstein condensates~\cite{UZ92,ORTR16,RCCP17}, and
quantum information theory~\cite{PS05,LORV05,BDV06,ODV08,WVVD12}. In particular, it has been shown
that the von Neumann entanglement entropy of its ground state grows logarithmically with the size
of the subsystem, as is the case for one-dimensional critical
systems~\cite{VLRK03,HLW94,Ko04,RM04} (although this model is actually not
critical~\cite{CFGRT16-LMG}).

As already noted in the original papers, the key to the solvability of the LMG model is the fact
that it can be mapped to a system of $N$ spin-$1/2$ particles with constant long-range
interactions of XY type in an external transverse magnetic field. In the isotropic (XX) case the
Hamiltonian of this effective model is a polynomial in~$\bJ^2$ and $J_z$, where $\bJ$ is the the
total spin operator, and can thus be exactly solved for arbitrary $N$. The general (non-isotropic)
LMG model can be solved in principle via the Bethe ansatz~\cite{PD99,MOPN06}, though in practice
this is less efficient than brute-force numerical diagonalization. In the thermodynamic limit,
however, the density of states of the latter model in the highest spin sector ($J=N/2$) has been
derived by means of a spin-coherent-state formalism~\cite{RVM07,RVM08}.

A wide family of models with long-range interactions of $\su(m)$ type generalizing the
\emph{isotropic} LMG model was recently introduced in Ref.~\cite{CFGRT16-LMG}. In analogy with the
latter model, the non-degenerate ground state of these novel models is given by a Dicke state
whose reduced density matrix for a subsystem of $L<N$ spins can be computed in closed form, which
in turn yields the entanglement entropy in the thermodynamic limit $N\to\infty$ with $L/N=\al$
finite. Although both the von Neumann and the Rényi entanglement entropies grow logarithmically
with the size~$L$ of the subsystem, the corresponding prefactor is independent of the Rényi
parameter, which implies that none of these models can be critical. Interestingly, for $m>3$ there
is at least one quantum phase whose Tsallis entanglement entropy~\cite{Ts88,Ts09} becomes
extensive for a suitable value of the Tsallis parameter. However, the full spectrum of these
models in general cannot be evaluated in closed form.

In this paper we introduce a family of generalized Lipkin--Meshkov--Glick (gLMG) models, with
interactions governed by an~$\su(m)$ integrable spin chain of Haldane--Shastry type. The latter
chains are the celebrated Haldane--Shastry (HS)~$\su(m)$ spin chain~\cite{Ha88,Sh88,HHTBP92},
which describes a circular array of equispaced spins with two-body long-range interactions
inversely proportional to the square of the (chord) distance, and its rational~\cite{Po93,Fr93}
and hyperbolic~\cite{FI94} analogues. Although the HS chain was originally introduced as a model
whose exact ground state coincides with Gutzwiller's variational wave function for the Hubbard
model in the limit of large on-site interaction~\cite{Gu63,GV87}, it soon proved of interest per
se in condensed matter and theoretical physics. Indeed, as pointed out by Haldane~\cite{Ha91b},
the spinon excitations of this chain provide one of the simplest examples of a quantum system
featuring fractional statistics (see also~\cite{HHTBP92,GS05,Gr09}). The HS chain is closely
connected to important conformal field theories like the $k=1$ Wess--Zumino--Novikov--Witten
model~\cite{Ha91b,BBS08}, and has recently been related to infinite matrix product
states~\cite{CS10}. Integrable extensions of the Haldane--Shastry chain with long-range
interactions involving more than two spins also play a key role for describing non-perturbatively
the spectrum of planar $\cN=4$ gauge theory in the context of the AdS-CFT
correspondence~\cite{BBL08,BBL09}. The interest in spin chains of HS type has been further
reinforced by recent developments in quantum simulation, as witnessed by the proposal of an
experimental realization of the HS chain using two internal atomic states of atoms trapped in a
photonic crystal waveguide~\cite{HGCK16}.

One of the key features of spin chains of Haldane--Shastry type is the fact that their partition
functions can be exactly computed for any number of spins~\cite{Po94,FG05,BFGR10} by exploiting
their connection with a corresponding spin dynamical model of Calogero--Sutherland
type~\cite{Ca71,Su71,Su72,In96} by means of a mechanism known as the Polychronakos ``freezing
trick''~\cite{Po94}. This has made it possible to check the validity of several fundamental
conjectures on the characterization of quantum chaos vs.~integrability~\cite{BT77,BGS84}. In
particular, it has been shown that spin chains of HS type do not behave as expected for a
`generic' integrable system, in the sense that the distribution of the spacings between
consecutive levels is not Poissonnian~\cite{FG05,BFGR08epl,BFGR10}.

The gLMG models that we introduce in this paper can also be regarded as a deformation of the
$\su(m)$ spin chains of HS type. More precisely, we add to the HS-type Hamiltonian a term
depending on the generators of the standard $\su(m)$ Cartan subalgebra, which commutes with the
former Hamiltonian. In particular, when this extra term is linear in the Cartan generators it can
be interpreted as an $\su(m)$ external magnetic field, and the corresponding models are the ones
studied in Ref.~\cite{EFG12}. Likewise, when the extra term is a suitable quadratic combination of
the Cartan generators we recover the models introduced in Ref.~\cite{CFGRT16-LMG}, which include
the isotropic~LMG model. We shall see that the Hilbert space of a general gLMG model decomposes as
a direct sum of subspaces with fixed magnon numbers, which are separately invariant under the
action of both the original HS-type Hamiltonian and the new term. By suitably adapting the
freezing trick, we shall be able to compute the partition function of the restriction of the
Hamiltonians of the three spin chains of HS type to the latter invariant subspaces. This in turn
yields the partition function of the full gLMG Hamiltonian, since the Cartan generators are
proportional to the identity on these subspaces. The knowledge of the partition function of the
gLMG models of HS type, as well as the restricted partition functions of the corresponding spin
chains, enables one to study several statistical properties of the spectrum of the latter models.
In particular, we have obtained strong numerical evidence that the level density of the
restriction of the HS-type chain Hamiltonians to subspaces with fixed magnon numbers follows a
Gaussian distribution in the large $N$ limit, as is known to be the case for the full spectrum of
these models~\cite{EFG09,EFG10}. We have also studied the distribution of the spacings between
consecutive levels of the restrictions of these models to the invariant subspaces, showing that it
follows the characteristic law for an approximately equispaced spectrum with normally distributed
energy levels~\cite{BFGR08epl,BFGR10}. Finally, we have numerically computed the thermodynamic
functions of gLMG models of HS type whose extra term is quadratic in the Cartan generators,
comparing them with the exact results for the original (HS-type) chains in the thermodynamic limit
derived in Ref.~\cite{EFG12}.

\section{The models}

The models we shall study in this paper are deformations of~$\su(m)$ spin chains with Hamiltonians
of the form
\begin{equation}\label{H0gLMG}
  H_0=\sum_{1\le i<j\le N} h_{ij}(1-\ep S_{ij})\,, \quad\ep=+,-\,,
\end{equation} 
with $h_{ij}\in\RR$. In the latter equation $S_{ij}$ is the operator permuting the $\su(m)$ spins
of the $i$-th and $j$-th particles, whose action on the canonical $\su(m)$ spin basis
\begin{equation}\label{spinb}
  \cS=\big\{\ket{s_1}\otimes\cdots\otimes\ket{s_N}\equiv\ket{s_1,\dots,s_N}\mid s_i=1,\dots,m\,,\
  1\le i\le N\big\}\,,
\end{equation}
is given by
\[
  S_{ij}\ket{s_1,\dots,s_i,\dots,s_j,\dots,s_N}=\ket{s_1,\dots,s_j,\dots,s_i,\dots,s_N}\,.
\]
These operators can be expressed in terms of the local (Hermitian) generators~$t^a_k$
($a=1,\dots,m^2-1$) of the fundamental representation of the $\su(m)$ algebra acting on the
$k$-th site (with the normalization $\tr(t^a_kt^b_k)=\frac12\,\de_{ab}$) as
\begin{equation}\label{Sijta}
S_{ij}=\frac1{m}+2\sum_{a=1}^{m^2-1}t^a_it^a_j\equiv\frac1{m}+2\,\bt_i\cdot\bt_j\,.
\end{equation}
We can thus write\footnote{Here and throughout the paper, all sums and products run from $1$ to
  $N$ unless otherwise specified.}
\[
H_0=-\ep\sum_{i\ne j}h_{ij}\,\bt_i\cdot\bt_j+E_0\,,
\]
with~$E_0=(1-\frac\ep m)\sum_{i<j}h_{ij}$. In particular, for~$m=2$ we
have~$\bt_k=\frac12\,\bsi_k$, where~$\bsi_k=(\si_k^1,\si_k^2,\si_k^3)$ are the three Pauli
matrices at the $k$-th site.

Let~$\cN_a$ denote the $a$-th magnon number operator defined by
\begin{equation}\label{Nadef}
  \cN_a\ket{s_1,\dots,s_N}=N_a\ket{s_1,\dots,s_N}\,,\qquad 1\le a\le m\,,
\end{equation}
where\footnote{We shall denote in what follows by~$|A|$ the cardinal of the set~$A$.}
\begin{equation}\label{Nask}
  N_a=\big|\big\{k=1,\dots,N\mid s_k=a\big\}\big|\,.
\end{equation}
The latter operators are related to the Hermitian generators of the standard Cartan subalgebra of
the Lie algebra~$\su(m)$, as we shall now explain. Indeed, let $J^a_k$ denote the operator whose
action on the Hilbert space of the $k$-th particle is given by
\begin{equation}\label{Jakdef}
  J_k^a\ket{s_k}=(\de_{a,s_k}-\de_{m,s_k})\ket{s_k}\,,\qquad 1\le a\le m-1\,.
\end{equation}
The $m-1$ commuting operators~$\iu J^a_k$ generate the standard Cartan subalgebra\footnote{This
  choice of the generators of the standard Cartan subalgebra of~$\su(m)$ is simply a matter of
  convenience. Note, however, that these generators are not orthogonal with respect to the usual
  Killing--Cartan scalar product, i.e., $\tr(J_k^aJ_k^b)\ne0$ for $a\ne b$.} of $\su(m)$ at each
site~$k$. We then define the global (Hermitian) Cartan generators
\[
  J^a\equiv\sum_{k=1}^N J_k^a\,,\qquad 1\le a\le m-1\,.
\]
From Eq.~\eqref{Jakdef} it then follows that
\[
J^a=\cN_a-\cN_m\,,\qquad 1\le a\le m-1\,.
\]
Summing over $a$ and taking into account that~$\sum_{a=1}^m\cN_a=N$ we obtain
\[
  \sum_{a=1}^{m-1}J^a=N-m\,\cN_m\,,\qquad 1\le a\le m-1\,.
\]
Using the last two equations we can express the magnon number operators in terms of the Cartan
subalgebra generators as
\begin{equation}\label{JaNa}
\cN_a=J^a(1-\de_{am})-\BJ+\frac Nm\,,\qquad 1\le a\le m\,,
\end{equation}
where
\[
\BJ\equiv\frac1m\sum_{a=1}^{m-1}J^a\,.
\]

We shall consider in what follows deformations $H=H_0+H_1$ of~\eqref{H0gLMG} in which
\begin{equation}
  \label{H1}
  H_1=h(\cN_1,\ldots,\cN_{m})
\end{equation}
is an analytic function of the magnon number operators~$\cN_a$. Note, first of all, that the
previous expression for~$H_1$ is not ambiguous, since $[\cN_a,\cN_b]=0$ for $1\le a,b\le m$. It is
also clear that $\iu H_1$ lies in the enveloping algebra of the $\su(m)$ Cartan subalgebra on
account of Eq.~\eqref{JaNa}. For this reason, we shall say that
\begin{eqnarray}\label{H}
  H=H_0+H_1&=\sum_{i<j} h_{ij}(1-\ep S_{ij})+h(\cN_1,\ldots,\cN_{m})\nonumber\\
   &=
     -\vep\sum_{i\ne j}h_{ij}\,\bt_i\cdot\bt_j+h(\cN_1,\ldots,\cN_{m})+E_0\,.
\end{eqnarray}
is an~$\su(m)$ \emph{generalized Lipkin--Meshkov--Glick} (gLMG) model. In particular,
when~$h_{ij}>0$ for all $i<j$, $\ep=+$ and $h$ is the quadratic polynomial
\[
  h(x_1,\dots,x_m)=\sum_{a=1}^{m-1}c_a(x_a-x_m-Nh_a)^2\,,\qquad \text{with}\quad h_a\in\RR\,,\
  c_a>0\,,
\]
we obtain the models whose ground state entanglement entropy was computed in closed form in
Ref.~\cite{CFGRT16-LMG}. The latter models include the original ($\su(2)$, isotropic) LMG model
when $h_{ij}=2/N$ for all $i<j$ and $c_1=1/(2N)$, up to a constant energy.

One of the fundamental properties of the Hamiltonian~\eqref{H} is that it preserves the subspaces
of the Hilbert space~$\cH\equiv(\CC^m)^{\otimes N}$ with a fixed magnon configuration. Indeed, let
us denote by~$\cH(\bN)$, where~$\bN=(N_1,\dots,N_m)$ and $|\bN|\equiv N_1+\cdots+N_m=N$, the
subspace of~$\cH$ whose elements are linear combinations of basis states
$\ket{s_1,\dots,s_N}\equiv\ket\bsv$ with magnon numbers $N_a$ (cf.~Eq.~\eqref{Nask}).
Clearly~$H_0$ leaves~$\cH(\bN)$ invariant, since each permutation operator~$S_{ij}$ does. On the
other hand, $\cN_a\ket s=N_a\ket s$ on~$\cH(\bN)$ by construction, and therefore
\[
  H_1=h(\bN)\quad \text{on }\cH(\bN)\,.
\]
Thus $H=H_0+H_1$ preserves~$\cH(\bN)$, as stated. It is also clear from the above discussion that
$[H_0,H_1]=0$, and that the eigenvalues of~$H^\bN\equiv H|_{\cH(\bN)}$ can be expressed as
\[
  E^0_i(\bN)+h(\bN)\,,\qquad 1\le i\le\dim\cH(\bN)\,,
\]
where~$\{E^0_i(\bN)\}_{1\le i\le\dim\cH(\bN)}$ is the spectrum of~$H_0^\bN\equiv H_0|_{\cH(\bN)}$.
Hence the partition function~$Z^{\bN}(T)$ of~$H^\bN$ is given by
\[
  Z^{\bN}(T)=q^{h(\bN)}\sum_{i=1}^{\dim\cH(\bN)}q^{E_i^0(\bN)}\equiv
  q^{h(\bN)}Z_0^{\bN}(T)\,,\qquad
  q\equiv\e^{-1/{k_\mathrm{B}T}}\,,
\]
where $Z_0^{\bN}(T)$ is the partition function of~$H_0^\bN$\,. Since
\[
  \cH=\bigoplus_{|\bN|=N}\cH(\bN)\,,
\]
the partition function of $H$ is given by
\begin{equation}\label{ZTfinal}
  Z(T)=\sum_{|\bN|=N}Z^\bN(T)=\sum_{|\bN|=N}q^{h(\bN)}Z_0^\bN(T)\,.
\end{equation}
Thus the partition function of the model~\eqref{H} is completely determined by the partition
functions $Z^\bN_0(T)$ of the restrictions of the spin chain Hamiltonian~$H_0$ to each of the
subspaces~$\cH(\bN)$. We shall see in the following sections that the latter partition functions
can be computed in closed form when $H_0$ is the Hamiltonian of one of the three spin chains of HS
type, namely the Haldane--Shastry~\cite{Ha88,Sh88}, Polychronakos--Frahm (PF)~\cite{Po93,Fr93} and
Frahm--Inozemtsev (FI)~\cite{FI94} chains. The chain sites of these integrable spin chains can be
expressed as
\begin{equation}\label{zs}
z_k=
\cases{
  k\pi/N\,,& for the HS chain\\
 \ze_k\,,& for the PF chain\\
  \e^{2\xi_k}\,,& for the FI chain\,,
}
\end{equation}
where $\ze_k$ and~$\xi_k$ respectively denote the $k$-th zero of the Hermite polynomial of
degree~$N$ and the generalized Laguerre polynomial $L_N^{\be-2N+1}$ with $\be>2(N-1)$. In all
three cases, the interaction strength is a function~$h_{ij}=h(z_i-z_j)$ of the
difference~$z_i-z_j$, namely
\begin{equation}\label{hs}
h(x)=
\cases{
  \tfrac12\sin^{-2}x\,,& for the HS chain\\
  x^{-2}\,,& for the PF chain\\
  \tfrac12\sinh^{-2}x\,,& for the FI chain\,.
}
\end{equation}
Remarkably, the (total) partition function~$Z_0(T)=\sum_{|\bN|=N}Z_0^{\bN}(T)$ of all of these
models can be computed in closed form by exploiting their close connection with their associated
spin Calogero--Sutherland models (see, e.g.,~\cite{Po94,FG05,BFGR10}). In the following sections
we shall adapt this technique, known in the literature as Polychronakos's freezing
trick~\cite{Po94}, to evaluate the restricted partition functions~$Z_0^\bN(T)$.

\section{The freezing trick}\label{sec.FT}

In this section we shall outline the computation of the restricted partition function~$Z_0^\bN$
for the Haldane--Shastry spin chain, which is the best known of these models and presents certain
technical subtleties stemming from its translation invariance. To this end, we first recall that
in this case~$H_0$ is related to the strong interaction limit of the spin Sutherland model
\[
  \HS=-\Delta+a\sum_{i\neq j}\sin^{-2}(x_i-x_j)(a-\ep S_{ij})\,,\qquad a>0\,,
\]
where~$\De\equiv\sum_i\pd ^2_{x_i}$. Indeed, we can write
\[
\HS=\Hsc+4a\Hsp(\bx)\,,
\]
where~$\bx=(x_1,\dots,x_N)$,
\[
\Hsc=-\Delta+a(a-1)\sum_{i\neq j}\sin^{-2}(x_i-x_j)
\]
is the scalar Sutherland model and
\[
\Hsp(\bx)=\frac12\sum_{i<j}\sin^{-2}(x_i-x_j)(1-\ep S_{ij})
\]
is obtained from $H_0$ replacing the chain sites~$z_i$ by the dynamical variables~$x_i$.
Since~$\HS$ and~$\Hsc$ are translation invariant, the total momentum is conserved and can be set
to zero by working in the center of mass frame. In the strong interaction limit~$a\to\infty$ the
eigenfunctions of~$\HS$ become sharply peaked at the coordinates of the minimum of the scalar
potential
\[
U(\bx)=\sum_{i\ne j}\sin^{-2}(x_i-x_j)
\]
in the configuration space ($A_{N-1}$ Weyl chamber)
\[
  A=\big\{\bx\in\RR^N\mid x_1<\cdots<x_N\big\}\,,
\]
which (up to an overall translation) coincide with the chain sites~$z_k=k\pi/N$\,. Thus, when
$a\gg1$ the eigenvalues of~$\HS$ are approximately given by
\[
E_{ij}\simeq E^{\mathrm{sc}}_i+4aE^0_j\,,\qquad a\gg1\,,
\]
where $E^{\mathrm{sc}}_i$ and $E^0_j$ respectively denote two arbitrary eigenvalues of~$\Hsc$
and~$H_0$. From the latter equation it immediately follows that the partition function~$Z_0(T)$ of
the Haldane--Shastry chain is given by the freezing trick formula
\begin{equation}\label{Z0FT}
  Z_0(T)=\lim_{a\to\infty}\frac{\ZS(4aT)}{\Zsc(4aT)}\,.
\end{equation}
This is the basis for the computation of~$Z_0(T)$ in Ref.~\cite{FG05}. We shall now show that
essentially the same procedure can be carried out to compute the restricted partition
functions~$Z_0^\bN(T)$. Essentially, this is due to the fact that the spin Hamiltonian~$\HS$
preserves the subspaces~$L^2(A)\otimes\cH(\bN)$ of its Hilbert space~$L^2(A)\otimes\cH$. Thus,
$Z_0^\bN$ can be obtained from the analogue of Eq.~\eqref{Z0FT}, namely
\begin{equation}\label{Z0NFT}
  Z_0^\bN(T)=\lim_{a\to\infty}\frac{\ZS^\bN(4aT)}{\Zsc(4aT)}\,,
\end{equation}
where $\ZS^\bN$ is the partition function of~$\HS^\bN=\HS|_{L^2(A)\otimes\cH(\bN)}$.

To begin with, note that the Hamiltonian~$\HS$ is equivalent to its symmetric/antisymmetric
extension to the Hilbert space~$\La_\pm\big(L^2(\RR^N)\otimes\cH\big)$, where~$\La_+$
(resp.~$\La_-$) is the symmetrizer (resp.~antisymmetrizer) with respect to permutations of the
particles' coordinates and~$\su(m)$ spin variables. This is basically due to the fact that any
point~$\bx\in\RR^N$ not lying on the singular hyperplanes~$x_i-x_j=0$ can be mapped in a unique
way to a point in~$A$ by a suitable permutation. As we shall see below, it shall be convenient for
what follows to identify~$\HS$ with its symmetric (resp.~antisymmetric) extension when~$\ep=1$
(resp.~$\ep=-1$). With this identification, it can be shown~\cite{FG05} that~$\HS$ is represented
by an upper triangular matrix in the appropriately ordered (non-orthonormal) basis with elements
\begin{equation}\label{psis}
\ket{\bp,\bsv}=\La_\ep\Big(\e^{2\iu\bp\cdot\bx}\prod_{i<j}\sin(x_i-x_j)^a\,\ket\bsv\Big)\,,
\end{equation}
where~$\ket\bsv\in\cS$ and~$\bp\equiv(p_1,\dots,p_N)\in\RR^N$ satisfy the following conditions:
\begin{enumerate}[i)]
\item The differences $n_i\equiv p_i-p_{i+1}$ ($1\leq i\leq N-1$) are nonnegative integers.
\item If $p_i=p_{i+1}$ then~$s_i\prec s_{i+1}$.
\item The total momentum of the state~$\ket{\bp,\bsv}$ vanishes, i.e., $\sum_i p_i=0$.
\end{enumerate}
In the second condition, the notation~$s_i\prec s_j$ stands for~$s_i<s_j$ when~$\ep=-1$ and
$s_i\le s_j$ when~$\ep=1$. The first condition is justified in Ref.~\cite{FG05}, the second one
can be arranged due to the symmetric/antisymmetric nature of the states~\eqref{psis}, while the
last one simply reflects that we are working in the center of mass frame. As shown in the latter
reference, the states $\ket{\bp,\bsv}$ should be ordered in such a way that $\ket{\bp,\bsv}$
precedes $\ket{\bp'\!,\bsv'}$ whenever $\bp<\bp'$, where the last notation means that $\bp$
precedes $\bp'$ in the lexicographic order. With this partial order, the action of~$\HS$ on the
basis~\eqref{psis} is upper triangular. More precisely~\cite{FG05},
\begin{equation}
  \label{Hsps}
  \HS\ket{\bp,\bsv}=E(\bp)\ket{\bp,\bsv}+\sum_{\mathclap{\bp'<\bp;\,\bsv'}}c(\bp',\bsv')\,\ket{\bp',\bsv'}\,,
\end{equation}
with~$c(\bp',\bsv')\in\CC$ and
\begin{equation}
  \label{Ep}
  E(\bp)=\sum_i\big[2p_i+a(N+1-2i)]^2\,.
\end{equation}
Since~$\HS$ preserves~$\cH(\bN)$, if the vector~$\bsv$ in Eq.~\eqref{Hsps} is such
that~$\ket\bsv\in\cH(\bN)$ then~$\ket{\bsv'}\in\cH(\bN)$ for all vectors~$\bsv'$ appearing in the
RHS of the latter equation. In other words, $\HS^\bN$ is also upper triangular with respect to the
basis~\eqref{psis}, where~$\ket\bsv\in\cS\cap\cH(\bN)$ and the quantum numbers~$(\bp,\bsv)$
satisfy conditions~i)--iii) above, ordered as previously explained. Moreover, by Eq.~\eqref{Hsps}
the eigenvalues of~$\HS^\bN$ are given by Eq.~\eqref{Ep}. Expanding the latter equation in powers
of $a$ we obtain
\[
E(\bp)=\Egs+4a\sum_ip_i(N+1-2i)+\Or(1)\,,
\]
where
\[
  \Egs=a^2\sum_i(N+1-2i)^2=\frac{a^2}3\,N(N^2-1)
\]
is the ground state energy of the ferromagnetic model~($\ep=1$). Thus in the limit $a\to\infty$ we
have
\[
\lim_{a\to\infty}q^{-\Egs/4a}\ZS^\bN(4aT)=\sum_{\bp,\bsv}q^{\sum_i p_i(N+1-2i)}\,,
\]
where the sum is extended to all~$(\bp,\bsv)$ satisfying conditions~i)--iii) above
with~$\ket\bsv\in\cS\cap\cH(\bN)$. Since the exponent is independent of the spin variables~$\bsv$,
the sum over~$\bsv$ can be immediately carried out, namely
\begin{equation}\label{ZSNsum}
\lim_{a\to\infty}q^{-\Egs/4a}\ZS^\bN(4aT)=\sum_{\bp}d(\bp,\bN,\ep)\,q^{\sum_i p_i(N+1-2i)}\,,
\end{equation}
where the spin degeneracy factor~$d(\bp,\bN,\ep)$ is the number of multiindices~$\bsv$ satisfying
condition~ii) above for a given~$\bp$ such that~$\ket\bsv\in\cS\cap\cH(\bN)$. In other words,
\begin{equation}
  \label{dpN}
  d(\bp,\bN,\ep)=\big|S(\bp,\ep)\cap\cH(\bN)\big|
\end{equation}
where
\begin{equation}\label{Spep}
  S(\bp,\ep)\equiv\left\{\ket{\bsv}\in\cS\mid p_i=p_{i+1}\Rightarrow s_{i}\prec
    s_{i+1}\,,\ 1\le i\le N-1\right\}\,.
\end{equation}
In order to evaluate the sum in Eq.~\eqref{ZSNsum}, we note that by conditions~i) and~iii) above
we can write the multiindex $\bp$ as
\begin{equation}\label{krho}
  \bp=(\overbrace{\rho_1,\dots,\rho_1}^{k_1},\dots,
  \overbrace{\rho_r,\dots,\rho_r}^{k_r})\,,
\end{equation}
with
\begin{equation}\label{condsrho}
  \fl k_1+\cdots+k_r=N,
  \quad k_1\rho_1+\cdots+k_r\rho_r=0\,,\quad \rho_i>\rho_{i+1},\quad
  \rho_i-\rho_{i+1}\in\NN\,.
\end{equation}
Thus the multiindex $\bp$ consists of $r$ \emph{blocks} of lengths~$k_1,\dots,k_r$. Calling
\begin{equation}
  \label{Kis}
  K_i=\sum_{j=1}^ik_j\,,\qquad 0\le i\le r\,,
\end{equation}
we have
\[
  \sum_i
  p_i(N+1-2i)=\sum_{i=1}^r\rho_i\quad\sum_{\mathclap{j=K_{i-1}+1}}^{K_i}(N+1-2j)=\sum_{i=1}^r\rho_ik_i(N-2K_i+k_i).
\]
Since~$d(\bp,\bN,\ep)$ obviously depends on~$\bp$ only through~$\bk\equiv(k_1,\dots,k_r)$, we can
rewrite Eq.~\eqref{ZSNsum} as
\begin{equation}
  \label{ZNSsimp}
  \fl
  \lim_{a\to\infty}q^{-\Egs/4a}\ZS^\bN(4aT)=\sum_{r=1}^N\sum_{\bk\in\cP^r_N}d(\bp,\bN,\ep)\;
  \sum_{\mathclap{\substack{\rho_1>\cdots>\rho_r,\,\rho_i-\rho_{i+1}\in\NN\cr k_1\rho_1+\cdots+k_r\rho_r=0}}}\;
  q^{\sum_{i=1}^r\rho_ik_i(N-2K_i+k_i)}\,,
\end{equation}
where~$\cP_N^r$ denotes the set of all partitions of the integer~$N$ in~$r$ parts with order taken
into account. The inner sum in Eq.~\eqref{ZNSsimp} was evaluated in Ref.~\cite{FG05}, with the
result
\begin{equation}\label{limainf}
  \sum_{\mathclap{\substack{\rho_1>\cdots>\rho_r,\,\rho_i-\rho_{i+1}\in\NN\cr k_1\rho_1+\cdots+k_r\rho_r=0}}}\;
  q^{\sum_{i=1}^r\rho_ik_i(N-2K_i+k_i)}=
  \prod_{i=1}^{r-1}\frac{q^{\cE(K_i)}}{1-q^{\cE(K_i)}}\,,
\end{equation}
where
\begin{equation}\label{dispHS}
  \cE(k)=k(N-k)\,.
\end{equation}
Substituting Eq.~\eqref{limainf} into Eq.~\eqref{ZNSsimp} we obtain
\begin{equation}
  \label{ZNS}
  \lim_{a\to\infty}q^{-\Egs/4a}\ZS^\bN(4aT)=\sum_{r=1}^N\en\sum_{\mathclap{\bk\in\cP^r_N}}d(\bp,\bN,\ep)
\prod_{i=1}^{r-1}\frac{q^{\cE(K_i)}}{1-q^{\cE(K_i)}}\,,
\end{equation}
where~$\bp$ is any multiindex of the form~\eqref{krho}. The partition function for the scalar
Hamiltonian was also evaluated in Ref.~\cite{FG05} in the large $a$ limit, namely
\begin{equation}\label{Zsclim}
  \lim_{a\to\infty}q^{-\Egs/4a}\Zsc(4aT)=\prod_{i=1}^N(1-q^{\cE(i)})^{-1}\,.
\end{equation}
Combining Eqs.~\eqref{ZNS}-\eqref{Zsclim} with Eq.~\eqref{Z0NFT} we finally obtain the following
explicit formula for the restricted partition function~$Z_0^\bN(T)$:
\begin{equation}
  \label{Z0NTexp}
  Z_0^\bN(T)=\sum_{r=1}^N\en\sum_{\mathclap{\bk\in\cP^r_N}}d(\bp,\bN,\ep)
  \prod_{i=1}^{r-1}q^{\cE(K_i)}\cdot\prod_{j=1}^{N-r}(1-q^{\cE(K'_j)})\,,
\end{equation}
where
\[
  \{K_1',\dots,K'_{N-r}\}=\{1,\dots,N-1\}-\{K_1,\dots,K_{r-1}\}
\]
and~$\bp$ is determined by~$\bk$ through Eq.~\eqref{krho}. Following a similar procedure for the
PF and FI chains we again obtain Eq.~\eqref{Z0NTexp}, but with~$\cE(k)$ in Eq.~\eqref{dispHS}
respectively given by~$k$ and~$k(\be-2N+k+1)$ (see Refs.~\cite{BFGR08epl,BFGR10} for more
details). In summary, the restricted partition function~$Z_0^\bN(T)$ for the three chains of HS
type is given by~Eq.~\eqref{Z0NTexp}, with \emph{dispersion relation}
\begin{equation}
  \label{dispHschains}
  \cE(k)=\cases{
    k(N-k)\,,& for the HS chain\\
    k\,,& for the PF chain\\
    k(\be-2N+k+1)\,,& for the FI chain\,.
  }
\end{equation}
Equations~\eqref{ZTfinal}-\eqref{Z0NTexp} yield an explicit formula for the partition function
of~the $\su(m)$~gLMG model~\eqref{H} with interactions $h_{ij}=h(z_i-z_j)$ given by
Eqs.~\eqref{zs}-\eqref{hs}, once the degeneracy factor~$d(\bp,\bN,\ep)$ is known.

\section{Degeneracy factor}
As we have seen in the previous section, in order to evaluate the partition function~$Z(T)$ of an
$\su(m)$ gLMG model of HS type through Eqs.~\eqref{ZTfinal}-\eqref{Z0NTexp}, we only need to
determine the degeneracy factor~$d(\bp,\bN,\ep)$ defined in Eq.~\eqref{dpN}. To this end, let us
fix~$\bp$ in Eq.~\eqref{krho} (with~$\bk\in\cP_N^r$) and take~$\bN=(N_1,\dots,N_m)$ such that
$N_a\in\NN_0\equiv\NN\cup\{0\}$ and~$|\bN|=N$. The degeneracy factor~$d(\bp,\bN,\ep)$ is obviously
much easier to compute in the antiferromagnetic case ($\ep=-1$), since by Pauli's principle the
$\su(m)$ spins in each block of length~$k_1,\dots,k_r$ in which the components of~$\bp$ are equal
must all be different (in fact, arranged in a strictly increasing sequence according to condition
ii) in the previous section).

\subsection{Anti-ferromagnetic case}
Let us define the vector $\brv=(r_1,\dots,r_m)$ by
\[
  r_i\equiv\big|\{j=1,\dots,r\mid k_j=i\}\big|\in\NN_0,\quad 1\le i\le m,
\]
so that
\begin{equation}\label{rm}
  r_1+\cdots+r_m=r\,,\qquad
  r_1+2r_2+\cdots+mr_m=N\,.
\end{equation}
In other words, $r_i$ is the number of blocks of length~$i$ in the expression~\eqref{krho}
for~$\bp$. Obviously $d(\bp,\bN,-)\equiv D_m(\brv,\bN)$, where~$D_m(\brv,\bN)$ denotes the number
of ways one can distribute $N_1$ spins $\ket{1}$, $N_2$ spins $\ket{2}$, \dots\,, $N_m$ spins
$\ket{m}$ in $r_1$ blocks of one site, $r_2$ blocks of two sites, \dots\,, $r_m$ blocks of $m$
sites, with all spins different in each block.

For $1\le i,j\le m$, let us denote by $N_{i,j}\in\NN_0$ the number of spins $\ket{i}$ in the $r_j$
blocks of $j$ sites, and define $\bN_i=(N_{i,1},\ldots,N_{i,m})$ such that $|\bN_i|=N_i$ for
$1\le i\le m$. We can find an expression for the degeneracy factor by counting the number of ways
one can fill the pattern of blocks so that all the spins in each block are different. To this end,
we start with an empty pattern and fill it as follows:
\begin{enumerate}[i)]
\item {\em Fill all the $r_m$ blocks of $m$ sites.}

  In the $r_m$ blocks of $m$ sites there must be $r_m$ spins of each type. We are left with
  $N_1-r_m$ spins $\ket{1}$, $N_2-r_m$ spins $\ket{2}$,$\dots$, $N_m-r_m$ spins $\ket{m}$ and a
  pattern of $r_1$ blocks of one site, $r_2$ blocks of two sites, $\dots$, $r_{m-1}$ blocks of
  $m-1$ sites.

\item {\em Distribute the remaining $N_m-r_m$ spins of type~$\ket{m}$ in the $r-r_m$ empty blocks
    left.}

  As in the previous step, we next fix a vector
  \[
    \bx=(x_1,\dots,x_{m-1})
  \]
  with $x_i\equiv N_{m,i}\in\NN_0$ and $|\bx|=N_m-r_m$. Clearly, the number of ways of
  distributing the $N_m-r_m$ spins $\ket m$ in the available $r-r_m$ blocks is given by the
  product of binomial coefficients $\prod_{i=1}^{m-1} \binom{r_i}{x_i}$.

\item {\em For each $\bx$ in step ii), we are left with a new pattern $\hatbrv\in\NN_0^{m-1}$ and
    new spins of types $1,\dots,m-1$ with magnon
    numbers~$(\hat N_1,\dots,\hat N_{m-1})\equiv\hat\bN$.}

  Remarkably, the new pattern $\hatbrv$ has no blocks of $m$ sites and the new vector $\hatbN$ has
  no spins $\ket{m}$. More precisely, for $i=1,\dots,m-1$ there are now $\hatrr_i=r_i-x_i+x_{i+1}$
  blocks of $i$ sites, i.e., the previous $r_i$ minus the occupied blocks of $i$ sites plus the
  occupied blocks of $i+1$ sites (note that we must take $x_m=0$, since all the blocks of~$m$
  sites were filled up in the first step). Thus, the new pattern $\hatbrv\equiv\hatbrv(\bx)$ and
  magnon vector $\hatbN$ are given by
  \begin{eqnarray}
    \fl\hatrr_i&=r_i-x_i+x_{i+1}\,,\en 1\le i\le m-2\,;\quad \hatrr_{m-1}=r_{m-1}-x_{m-1},
                 \label{rbarvec}\\
    \fl\hatN_i&=N_i-r_m\,,\en1\le i\le m-1,\label{Nbarvec}
  \end{eqnarray}
  and therefore
  \begin{equation}\label{DmrN}
    D_m(\brv,\bN)=\sum_{\mathclap{|\bx|=N_m-r_m}}\kern1.25em\prod_{i=1}^{m-1}
    \binom{r_i}{x_i}\cdot D_{m-1}\big(\hatbrv(\bx),\hatbN\big)\,.
  \end{equation}
  Note that the new vectors~$\hatbN$ and~$\hatbrv$ satisfy a relation analogous to the last
Eq.~\eqref{rm}, namely (by Eqs.~\eqref{rbarvec}-\eqref{Nbarvec})
  \begin{equation}
    |\hatbN|=N-N_m-(m-1)r_m
    =\hatrr_1+2\,\hatrr_2+\cdots+(m-1)\,\hatrr_{m-1}\,.
    \label{Nbar}
  \end{equation}
\item {\em Iterate the process described above.}
  
  By Eq.~\eqref{DmrN}, we can express the degeneracy factor
  \[
    D_m(\brv,\bN)\equiv D_m(\brv^{(m)},\bN^{(m)})
  \]
  as a linear combination of degeneracy factors
  \[
    D_{m-1}(\hatbrv,\hatbN)\equiv D_{m-1}(\brv^{(m-1)},\bN^{(m-1)})\,.
  \]
  This process can be iterated, by expressing each term
  $D_{m-1}\left(\brv^{(m-1)},\bN^{(m-1)}\right)$ in Eq.~\eqref{DmrN} in terms of degeneracy
  factors
  \[
    D_{m-2}\left((\brv^{(m-1)})\hat{\vphantom{\brv^m}},
      (\bN^{(m-1)})\hat{\vphantom{\hatbrv^m}}\;\right)\equiv
    D_{m-2}(\brv^{(m-2)},\bN^{(m-2)})\,,
  \]
  and so on. We thus obtain the recursion relation
  \begin{equation}\label{recrel}
    \fl D_k(\brv^{(k)},\bN^{(k)})=\sum_{\mathclap{|\bx|=N_k^{(k)}-r_k^{(k)}}}\kern1.6em
    \prod_{i=1}^{k-1}{{r_i^{(k)}}\choose{x_i}}\cdot
    D_{k-1}\big(\brv^{(k-1)}(\bx),\bN^{(k-1)}\big),
  \end{equation}
  where
  \[
    \brv^{(k)},\bN^{(k)}\in\NN_0^k\,;\quad \bx, \brv^{(k-1)}(\bx),\bN^{(k-1)}\in\NN_0^{k-1}\,,
  \]
  with
  \[
    r_i^{(k-1)}(\bx)=r_i^{(k)}-x_i+x_{i+1}\en(x_k\equiv0)\,,\quad
    N_i^{(k-1)}=N_i^{(k)}-r_k^{(k)}
  \]
  and~$\brv^{(m)}\equiv\brv$, $\bN^{(m)}\equiv\bN$. The above recursion relation, together with
  the obvious initial condition~$D_1=1$, fully determines~$D_m(\brv,\bN)$.
\end{enumerate}
In Section~\ref{sec.exa} we shall illustrate the above procedure for computing the degeneracy
factor~$d(\bp,\bN,-)\equiv D_m(\brv,\bN)$ with several examples. Once~$D_m$ is determined, the
restricted partition $Z_0^{\bN,(-)}(T)$ in the antiferromagnetic case is obtained from
Eq.~\eqref{Z0NTexp}, namely
  \begin{equation}
    \label{Z0Naf}
    Z_0^{\bN,(-)}(T)=\sum_{\mathclap{r=\lceil N/m\rceil}}^N\kern1.4em\sum_{\mathclap{\bk\in\cP^r_N}}D_m(\brv,\bN)
    \prod_{i=1}^{r-1}q^{\cE(K_i)}\cdot\prod_{j=1}^{N-r}(1-q^{\cE(K'_j)})\,,
\end{equation}
where the range of the last sum comes from condition~ii) above, since in the antiferromagnetic case
the lengths~$k_i$ of the blocks in Eq.~\eqref{krho} are all at most equal to~$m$. The partition
function of the corresponding gLMG model of HS type~\eqref{H} can then be computed from Eq.~\eqref{ZTfinal},
with the result
\begin{equation}
  \label{ZTaf}
  \fl
  Z^{(-)}(T)=\sum_{|\bN|=N}q^{h(\bN)}\sum_{\mathclap{r=\lceil N/m\rceil}}^N\kern1.4em\sum_{\mathclap{\bk\in\cP^r_N}}D_m(\brv,\bN)
  \prod_{i=1}^{r-1}q^{\cE(K_i)}\cdot\prod_{j=1}^{N-r}(1-q^{\cE(K'_j)})\,.
\end{equation}

\subsection{Ferromagnetic case}

A similar procedure could be followed in principle to compute the degeneracy factor~$d(\bp,\bN,+)$
in the ferromagnetic case~$\ep=1$. The main difference is that now each value of the~$\su(m)$ spin
can be used more than once to fill the blocks of length $k_1,\dots,k_r$ determined by the
multiindex~$\bp$ in Eq.~\eqref{krho}, which considerably complicates matters.

In practice, it is much easier to derive the ferromagnetic partition function~$Z^{(+)}$ from the
antiferromagnetic one~$Z^{\mathrm(-)}$ computed in the previous subsection by means of the
identity
\begin{equation}
  \label{HFAF}
  H_0^{(+)}+H_0^{(-)}=\sum_{i\ne j}h_{ij}\equiv \Emax(N)\,,
\end{equation}
where~$H_0^{(\pm)}$ denotes the Hamiltonian~\eqref{H0gLMG} with~$\ep=\pm$. The
constant~$\Emax(N)$, which is the maximum energy of~$H_0^{(-)}$, can be easily computed in closed
form for each of the interactions~\eqref{hs}-\eqref{zs} taking into account the
identity~\cite{BBH10}
\begin{equation}\label{Emax}
  \Emax(N)=\sum_{i=1}^{N-1}\cE(i)\,,
\end{equation}
namely
\[
\Emax=\cases{\frac{N}6\,(N^2-1)\,,&for the HS chain\\
\frac{N}2\,(N-1)\,,& for the PF chain\\
\frac{N}6\,(N-1)(3\be-4N+2)\,,& for the FI chain\,.}
\]
From Eq.~\eqref{HFAF} it immediately follows that the restricted partition
functions~$Z_0^{\bN,(\pm)}$ of~$H_0^{(\pm)}$ are related by
\[
Z_0^{\bN,(+)}(q)=q^{\Emax(N)}Z_0^{\bN,(-)}(q^{-1})\,.
\]
Using Eqs.~\eqref{Z0NTexp} and~\eqref{Emax} we easily obtain
\begin{eqnarray}
  \fl
  Z_0^{\bN,(+)}(T)&=q^{\Emax(N)}\sum_{\mathclap{r=\lceil
                    N/m\rceil}}^N\kern1.4em\sum_{\mathclap{\bk\in\cP^r_N}}
                    D_m(\brv,\bN)
                    \prod_{i=1}^{r-1}q^{-\cE(K_i)}\cdot\prod_{j=1}^{N-r}(1-q^{-\cE(K'_j)})\nonumber\\
  \fl&=\sum_{\mathclap{r=\lceil
       N/m\rceil}}^N(-1)^{N-r}
       \sum_{\mathclap{\bk\in\cP^r_N}}D_m(\brv,\bN)
       \prod_{i=1}^{N-r}(1-q^{\cE(K'_i)})\,,
  \label{Z0NTferro}
\end{eqnarray}
where~$D(\brv,\bN)$ is the antiferromagnetic degeneracy factor computed in the previous
subsection. By Eq.~\eqref{ZTfinal}, the partition function of the ferromagnetic
Hamiltonian~$H^{(+)}\equiv H_0^{(+)}+H_1$ is given by
\begin{equation}
  Z^{(+)}(T)=\sum_{\mathclap{|\bN|=N}}q^{h(\bN)}\sum_{\mathclap{r=\lceil
       N/m\rceil}}^N(-1)^{N-r}\sum_{\mathclap{\bk\in\cP^r_N}}D_m(\brv,\bN)
  \prod_{i=1}^{N-r}(1-q^{\cE(K'_i)})\,.
\label{Zferro}
\end{equation}

\section{Examples}\label{sec.exa}

\subsection{$\su(2)$}

In this case~$\brv=(r_1,r_2)$, $\bN=(N_1,N_2)$, and the recursion relation~\eqref{recrel} with
$D_1=1$ immediately yields
\[
  D_2(\brv,\bN)=\binom{r_1}{N_2-r_2}\,.
\]
Expressing~$r_1,r_2$ in terms of~$r$ and~$N$ by means of the relations~$r=r_1+r_2$,
$N=N_1+N_2=r_1+2\,r_2$ we finally obtain
\[
  D_2(\brv,\bN)={{2r-N}\choose{r-N_1}}={{2r-N}\choose{r-N_2}}\,.
\]
Thus the restricted partition function of the $\su(2)$ chains~\eqref{H0gLMG} of HS type is given
by
\[
  Z_0^{\bN}(T)=\sum_{r=1}^N(-1)^{\frac{1+\ep}2(N-r)}\sum_{\mathclap{\bk\in\cP^r_N}}\binom{2r-N}{r-N_1}q^{\frac{1-\ep}2\sum\limits_{i=1}^{r-1}\cE(K_i)}
  \prod_{i=1}^{N-r}(1-q^{\cE(K'_i)})\,.
\]
By Eq.~\eqref{ZTfinal}, the partition function of the corresponding~$\su(2)$ gLMG model reads
\[
  \fl
  Z(T)=\sum_{N_1=0}^Nq^{h(N_1,N-N_1)}\sum_{r=1}^N(-1)^{\frac{1+\ep}2(N-r)}\sum_{\mathclap{\bk\in\cP^r_N}}
  \binom{2r-N}{r-N_1}q^{\frac{1-\ep}2\sum\limits_{i=1}^{r-1}\cE(K_i)}
  \prod_{i=1}^{N-r}(1-q^{\cE(K'_i)})\,.
\]

\subsection{$\su(3)$}
Let $\brv=(r_1,r_2,r_3)$ and $\bN=(N_1,N_2,N_3)$ such that $r=r_1+r_2+r_3$, $N=N_1+N_2+N_3$ and
$r_1+2\,r_2+3\,r_3=N$. We then have
\begin{eqnarray}
  \fl D_3(\brv,\bN)
  &=\!\!\sum_{x_1+x_2=N_3-r_3}\!\!{{r_1}\choose{x_1}}{{r_2}\choose{x_2}}D_2(\hatbrv,\hatbN)
    =\!\!\sum_{x_1+x_2=N_3-r_3}\!\!{{r_1}\choose{x_1}}{{r_2}\choose{x_2}}{{2\,\hatrr-\hatN}%
    \choose{\hatrr-\hatN_2}}\nonumber\\
  \fl &=\!\!\sum_{x_1+x_2=N_3-r_3}\!\!{{r_1}\choose{x_1}}{{r_2}\choose{x_2}}
        {{2r-2x_1-N+N_3}\choose{r-x_1-N_2}},\label{D3}
\end{eqnarray}
where we have used the identities $\hatrr=r-r_3-x_1$, $\hatN=N-N_3-2\,r_3$ and $\hatN_2=N_2-r_3$.

\subsection{$\su(4)$}
Let $\brv=(r_1,r_2,r_3,r_4)$ and $\bN=(N_1,N_2,N_3,N_4)$ such that $r=r_1+r_2+r_3+r_4$,
$N=N_1+N_2+N_3+N_4=r_1+2\,r_2+3\,r_3+4\,r_4$. Using Eq.~\eqref{D3}
with~$(\hatbrv,\hatN)\equiv(\brv^{(3)},\bN^{(3)})$ in place of~$(\brv,\bN)$ we easily obtain
\begin{eqnarray}
  \fl  D_4(\brv,\bN)&=\!\!\!\sum_{x_1+x_2+x_3=N_4-r_4}
                      \!\!\!{{r_1}\choose{x_1}}{{r_2}\choose{x_2}}{{r_3}\choose{x_3}}D_3(\hatbrv,\hatbN)\nonumber\\
  \fl &=\sum\limits_{\mathclap{x_1+x_2+x_3=N_4-r_4}}{\textstyle
        {{r_1}\choose{x_1}}{{r_2}\choose{x_2}}{{r_3}\choose{x_3}}\kern .5em
        }\sum\limits_{\mathclap{y_1+y_2=N_3-r_4-r_3+x_3}}{\textstyle
        \binom{r_1-x_1+x_2}{y_1}\binom{r_2-x_2+x_3}{y_2}
        \binom{2\,r-2\,x_1-2\,y_1-N_1-\,N_2}{r-x_1-y_1-N_2}
        }.
                \label{D4}
\end{eqnarray}

\section{The LMG-PF model}\label{sec.LMG-PF}
When~$H_0$ is the Hamiltonian of the PF chain the restricted partition function~$Z_0^\bN(T)$, and
hence the partition function~$Z(T)$ of the corresponding LMG-PF model~\eqref{H}, can be
considerably simplified. Indeed, in this case
\begin{equation}\label{HScal}
  \HS=-\Delta+a r^2+\sum_{i\ne j}\frac{a(a-\ep S_{ij})}{(x_i-x_j)^2}
  =\Hsc+2a\hat H_0(\bx)\,,
\end{equation}
where $r^2\equiv\sum_ix_i^2$,
\[
  \Hsc=-\Delta+a r^2+\sum_{i\ne j}\frac{a(a-1)}{(x_i-x_j)^2}
\]
is the scalar Calogero model and
\[
\hat H_0(\bx)=\sum_{i<j}\frac{1-\ep S_{ij}}{(x_i-x_j)^2}
\]
is obtained from $H_0$ by the formal substitution~$z_i\mapsto x_i$. Proceeding as in
Section~\ref{sec.FT} we obtain the analogue of Eq.~\eqref{Z0NFT}, namely
\begin{equation}\label{Z0NPF}
  Z_0^\bN(T)=\lim_{a\to\infty}\frac{\ZS^\bN(2aT)}{\Zsc(2aT)}\,,
\end{equation}
where the partition function~$\Zsc(2aT)$ of the scalar Calogero model is given by~\cite{BFGR08epl}
\begin{equation}\label{ZscCal}
  \Zsc(2aT)=q^{\Egs/2a}\prod_i(1-q^i)^{-1}\,,\quad \Egs\equiv a^2N(N-1)+aN\label{EpCal}\,.
\end{equation}
In order to compute~$\ZS^\bN(2aT)$, we note~\cite{BFGR08epl} that the Hamiltonian~\eqref{HScal} of the spin
Calogero model is upper triangular in the basis with elements
\begin{equation}\label{ketpsPF}
\ket{\bp,\bsv}=\e^{-ar^2/2}\prod_{i<j}|x_i-x_j|^a\La_\ep\bigg(\prod_ix_i^{p_i}\ket\bsv\bigg)\
\end{equation}
partially ordered by the total degree~$|\bp|$, with corresponding eigenvalues
\begin{equation}
E(\bp)=2a(p_1+\cdots+p_N)+\Egs\,.
\end{equation}
Of course, we must choose the quantum numbers~$(\bp,\bsv)$ in such a way that the
states~\eqref{ketpsPF} are actually a basis. The main difference with the HS and FI models is that
only in this case~$E(\bp)$ and the admissible partial order of the basis states~\eqref{ketpsPF} do
not depend on the ordering of the components of~$\bp$~\cite{FG05,BFGR08epl,BFGR10}. As a
consequence, we can choose the quantum numbers~$(\bp,\bsv)$ in each
subspace~$\La_\ep\big[L^2(\RR^N)\otimes\cH(\bN)\big]$ as follows:
\begin{enumerate}[i)]
\item We first order the components of the spin quantum number~$\bsv$ increasingly, so that
\[
  \bsv=\big(\overbrace{1,\dots,1}^{N_1},\dots,\overbrace{m,\dots,m}^{N_m}\big)
\]
is now fixed.
\item In each block of~$\bsv$ with fixed magnon number~$\ket a$ we order the corresponding
  components of the vector~$\bp$ also increasingly, so that $\bp=(\brho^1,\dots,\brho^m)$ with
\[
  \brho^j\equiv(\rho^j_1,\dots,\rho^j_{N_j})
\]
and $\rho^j_i\,\prec\,\rho^j_{i+1}$.
\end{enumerate}
We thus have
\[
  E(\bp)=\Egs+2a\sum_{i=1}^Np_i=\Egs+2a\sum_{j=1}^m\sum_{i=1}^{N_j}\rho^j_i\,,
\]
and therefore
\[
  q^{-\Egs/2a}\ZS^{\bN}(2aT)=\sum_{\rho^k_i\prec\rho^k_{i+1}}\prod_{j=1}^mq^{\rho^j_1+\cdots+\rho^j_{N_j}}
  =\prod_{j=1}^m\kern
  1.75em\sum_{\mathclap{0\le\rho^j_1\prec\,\cdots\,\prec\rho^j_{N_j}}}q^{\rho^j_1+\cdots+\rho^j_{N_j}}\,.
\]
The inner sum in the latter formula can be computed in closed form, with the result
\[
\sum_{\mathclap{0\le\rho^j_1\prec\cdots\prec\rho^j_{N_j}}}q^{\rho^j_1+\cdots+\rho^j_{N_j}}=
q^{\frac{1-\ep}4N_j(N_j-1)}\prod_{i=1}^{N_j}(1-q^i)^{-1}\equiv q^{\frac{1-\ep}4N_j(N_j-1)}{(q)}_{N_j}^{-1}
\,,
\]
and thus
\[
  \ZS^\bN(2aT)=\prod_{j=1}^mq^{\frac{1-\ep}4N_j(N_j-1)}{(q)}_{N_j}^{-1}=q^{\frac{1-\ep}4\sum\limits_{j=1}^mN_j(N_j-1)}
  \prod_{j=1}^m{(q)}_{N_j}^{-1}\,.
\]
From this equation and Eqs.~\eqref{Z0NPF}-\eqref{ZscCal} we obtain the following closed-form
expression for the restricted partition function of the PF chain:
\[
  Z^{\bN}_0(T)=q^{\frac{1-\ep}4\sum\limits_{j=1}^mN_j(N_j-1)}\frac{(q)_N}{\prod_{j=1}^m(q)_{N_j}}\,.
\]
Finally, by Eq.~\eqref{ZTfinal} the partition function of the LMG-PF model is given by
\begin{eqnarray}
  Z(T)&=\quad\sum_{\mathclap{N_1+\cdots+N_m=N}}q^{h(\bN)+\frac{1-\ep}4\smash{\sum\limits_{j=1}^mN_j(N_j-1)}}
  \frac{(q)_N}{\prod_{j=1}^m(q)_{N_j}}\nonumber\\&\equiv\quad
  \sum_{\mathclap{N_1+\cdots+N_m=N}}q^{h(\bN)+\frac{1-\ep}4\smash{\sum\limits_{j=1}^mN_j(N_j-1)}}
                                                   \qbinom{N}{N_1,\dots,N_m}q\,.
                                                   \label{ZTPF}
\end{eqnarray}
In particular, for~$h=0$ we recover the well-known formula for the partition function of the PF
chain in Ref.~\cite{Po94}.

\section{Analysis of the spectrum and thermodynamics}

\begin{figure}[t!]
  \includegraphics[height=.31\textwidth]{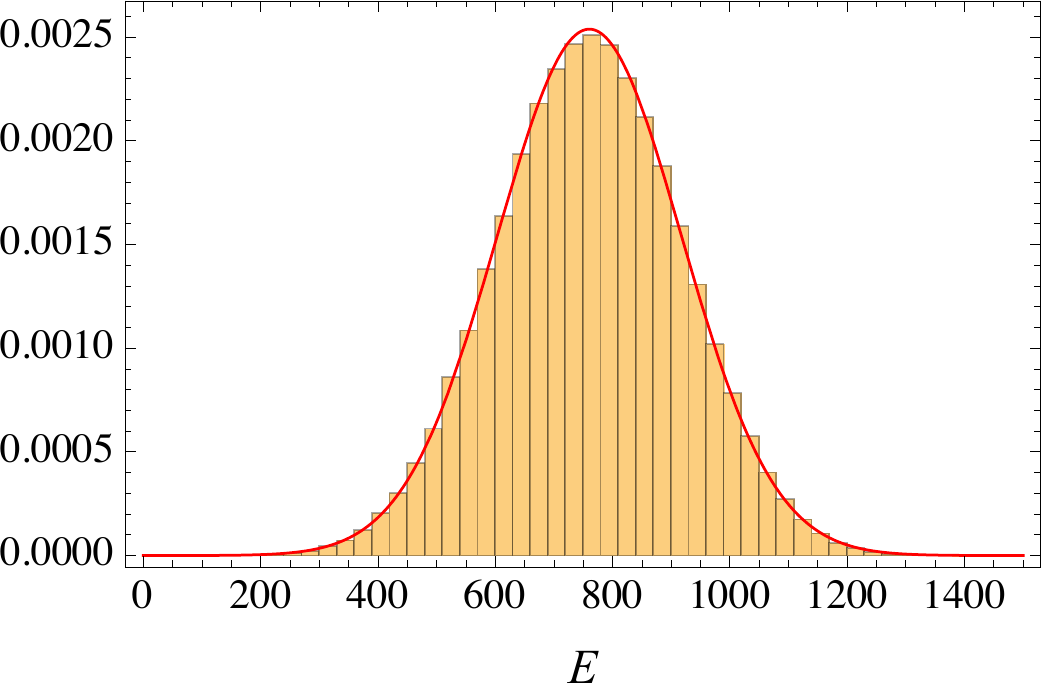}\hfill
  \includegraphics[height=.31\textwidth]{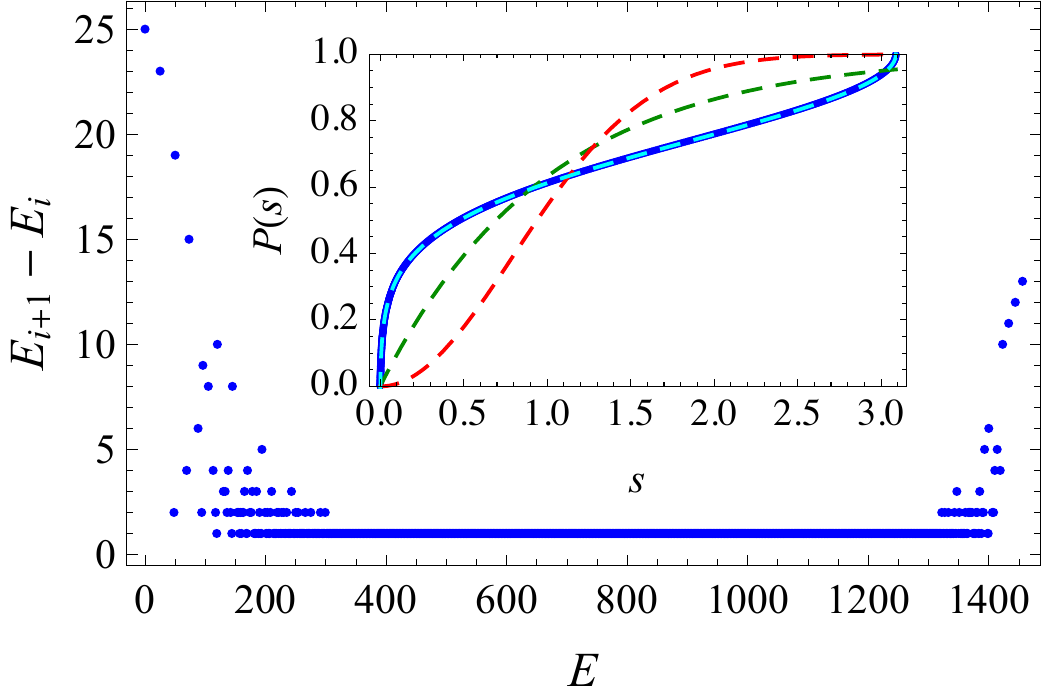}\\[2mm]
  \includegraphics[height=.31\textwidth]{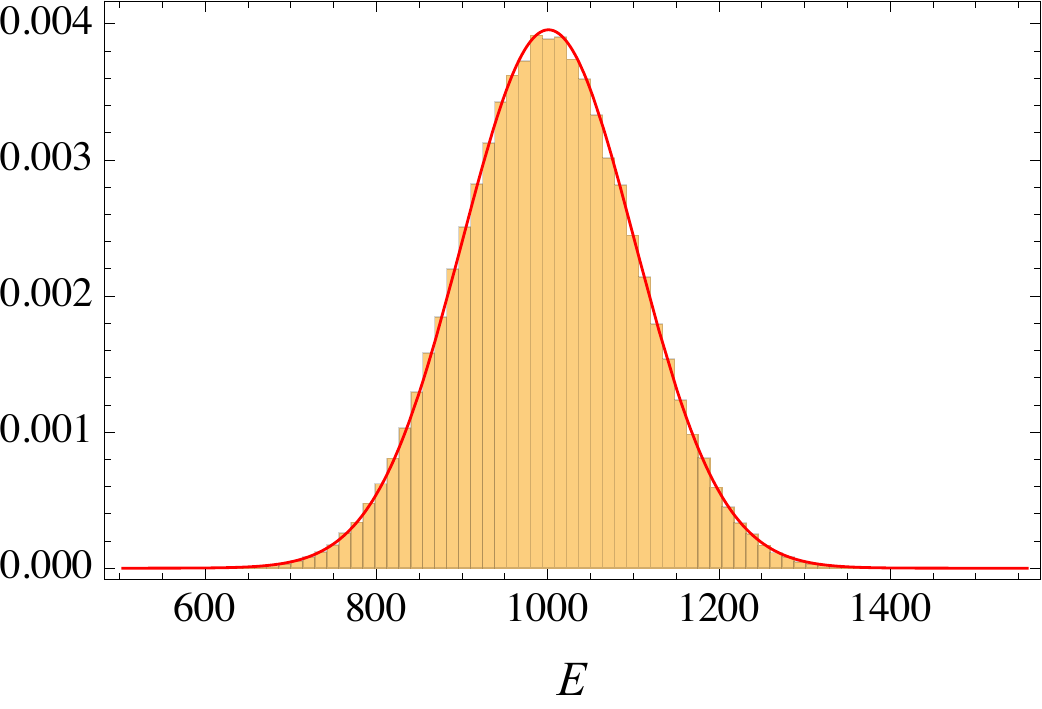}\hfill
  \includegraphics[height=.31\textwidth]{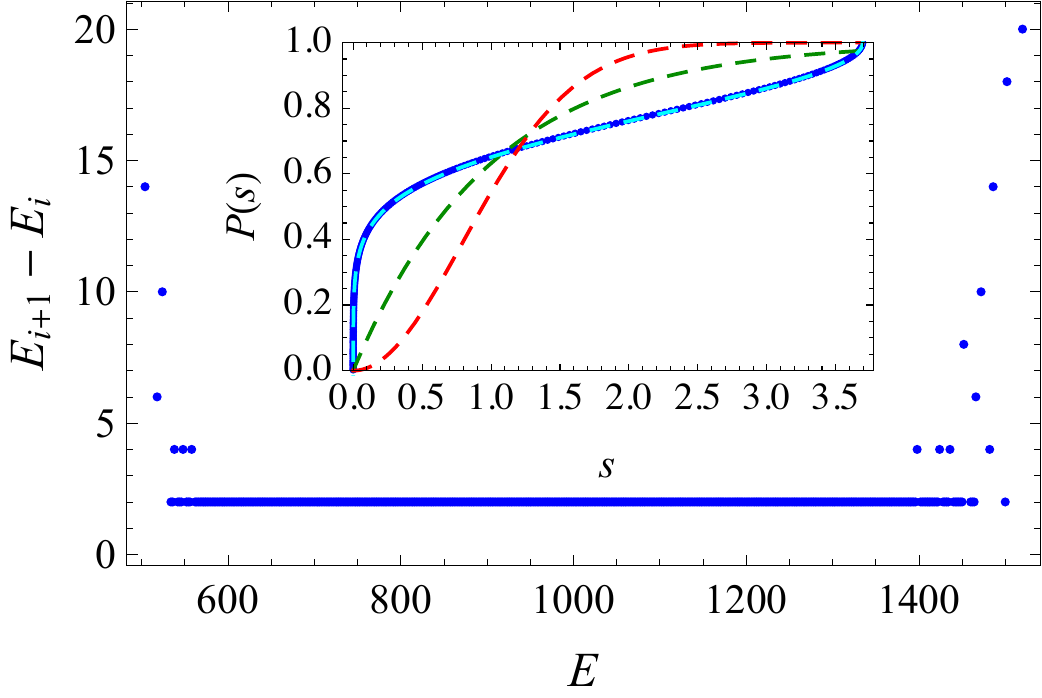}
  \caption{Left: level density histogram (normalized to unity) for the HS chain in a subspace with
    fixed magnon numbers $N_1=N_2=13$ ($\su(2)$, top) and $N_1=N_2=N_3=8$ ($\su(3)$, bottom)
    compared to a Gaussian distribution (continuous red line). Right: differences between
    consecutive levels of the raw spectrum (main plots) and cumulative spacings distribution of
    the unfolded spectrum (insets) in the latter configurations. The dashed red, green and light
    blue curves in the insets are respectively the Wigner (GOE) and Poisson cumulative
    distributions and the law~\eqref{Ps}. (In all cases, we have used natural units $\hbar=2M=1$.)}
  \label{fig.diffsspacings}
\end{figure}

In this section we shall take advantage of the knowledge of the restricted partition function of
the gLMG models~\eqref{H} to study several statistical properties of their spectrum and analyze
the behavior of their thermodynamic functions for large $N$. To begin with, we have examined the
level density of the restriction of the Hamiltonian to subspaces with a fixed magnum content.
Since $H_1$ is constant on these subspaces, this is of course equivalent to studying the level
density of the corresponding spin chains of HS type. It is well-known in this
respect~\cite{EFG09,EFG10} that the level density of the {\em complete} spectrum of the latter
models becomes normally distributed in the $N\to\infty$ limit, essentially due to the existence of
a description of the spectrum in terms of Haldane's motifs~\cite{HHTBP92,BBH10}. We have computed
the spectrum of the HS chain for up to $N=26$ for $\su(2)$ and $N=24$ for $\su(3)$ in the largest
subspace~$\cH(\bN)$ (with $N_i=N/m$ for all~$i$). Our results clearly indicate that the spectrum
of the restriction of~$H_0$ to this subspace is also normally distributed (see
Fig.~\ref{fig.diffsspacings}, left), with parameters $\mu$ and $\si$ given by the mean and
standard deviation of the restricted spectrum. For the FI and PF chains we have obtained similar
results. This fact suggests~\cite{EFG10} that in all three cases there might be a formula for the
energies in each sector of the spectrum with fixed magnon numbers in terms of motifs.

Since the continuous part of the cumulative level density in each sector can be well approximated
by a Gaussian distribution, the energies of the ``unfolded'' spectrum~\cite{Ha01} can be taken as
\[
\eta_i=\int_{-\infty}^{E_i}g(t)\diff t\,,\qquad
g(E)=\frac1{\sqrt{2\pi}\si}\,\e^{-\frac{(E-\mu)^2}{2\si^2}}\,,\quad
i=1,\ldots,n\,.
\]
According to a long-standing conjecture due to Berry and Tabor~\cite{BT77}, the distribution of
the (normalized) spacings between consecutive levels of the unfolded spectrum, defined as
\[
s_i=(n-1)\frac{\eta_{i+1}-\eta_i}{\eta_n-\eta_1}\,,\quad i=1,\ldots,n-1,
\]
is expected to be Poissonian for `generic' integrable systems. On the other hand, for a chaotic
system the well-known Bohigas--Giannoni--Schmit conjecture posits that this distribution should be
given by the Wigner distribution corresponding to the appropriate ensemble of random
matrices~\cite{BGS84}. In Refs.~\cite{FG05,BFGR10,BFGR08epl} it was observed that the distribution
$p(s)$ of the spacings between consecutive levels of the {\em whole} spectrum of all three chains
of HS type follows none of the above distributions, but is typically given by the `square root of
a logarithm' law
\begin{equation}\label{Ps}
  P(s)=1-\frac2{\sqrt\pi\,s_{\mathrm{max}}}\sqrt{\log\biggl(\frac{s_{\mathrm{max}}}s\biggr)}\,,
\end{equation}
where~$P(s)=\int_0^sp(s')\diff s'$ is the cumulative distribution and~$s_{\mathrm{max}}$ is the
maximum spacing. As shown in Refs.~\cite{BFGR08,BFGR09}, this is due to the fact that the raw
spectrum of the latter chains is approximately equispaced and normally distributed. We have
computed the distribution of consecutive (normalized) spacings in the subspaces mentioned above
for the HS, PF and FI chains for $m=2$ and $3$. In all cases, the cumulative spacings
distribution~$P(s)$ fits Eq.~\eqref{Ps} with remarkable accuracy (see the insets
Fig.~\ref{fig.diffsspacings}, right, for the HS chain). This clearly suggests that the (raw)
spectrum of the restriction of the three HS-type chains to subspaces with fixed magnon content is
also approximately equispaced. We have also verified that this conclusion is indeed correct for
all three chains of HS type. For instance, for the $\su(2)$ HS chain with $N_1=N_2=13$
(cf.~Fig.~\ref{fig.diffsspacings}, top right) $93.8\%$ of the spacings between consecutive levels
of the raw spectrum is equal to $1$, while for the $\su(3)$ HS chain with $N_1=N_2=N_3=8$
(cf.~Fig.~\ref{fig.diffsspacings}, bottom right) the predominant spacing is again~$1$ and occurs
$95.7\%$ of the times.

\begin{figure}[t!]
  \includegraphics[height=.31\textwidth]{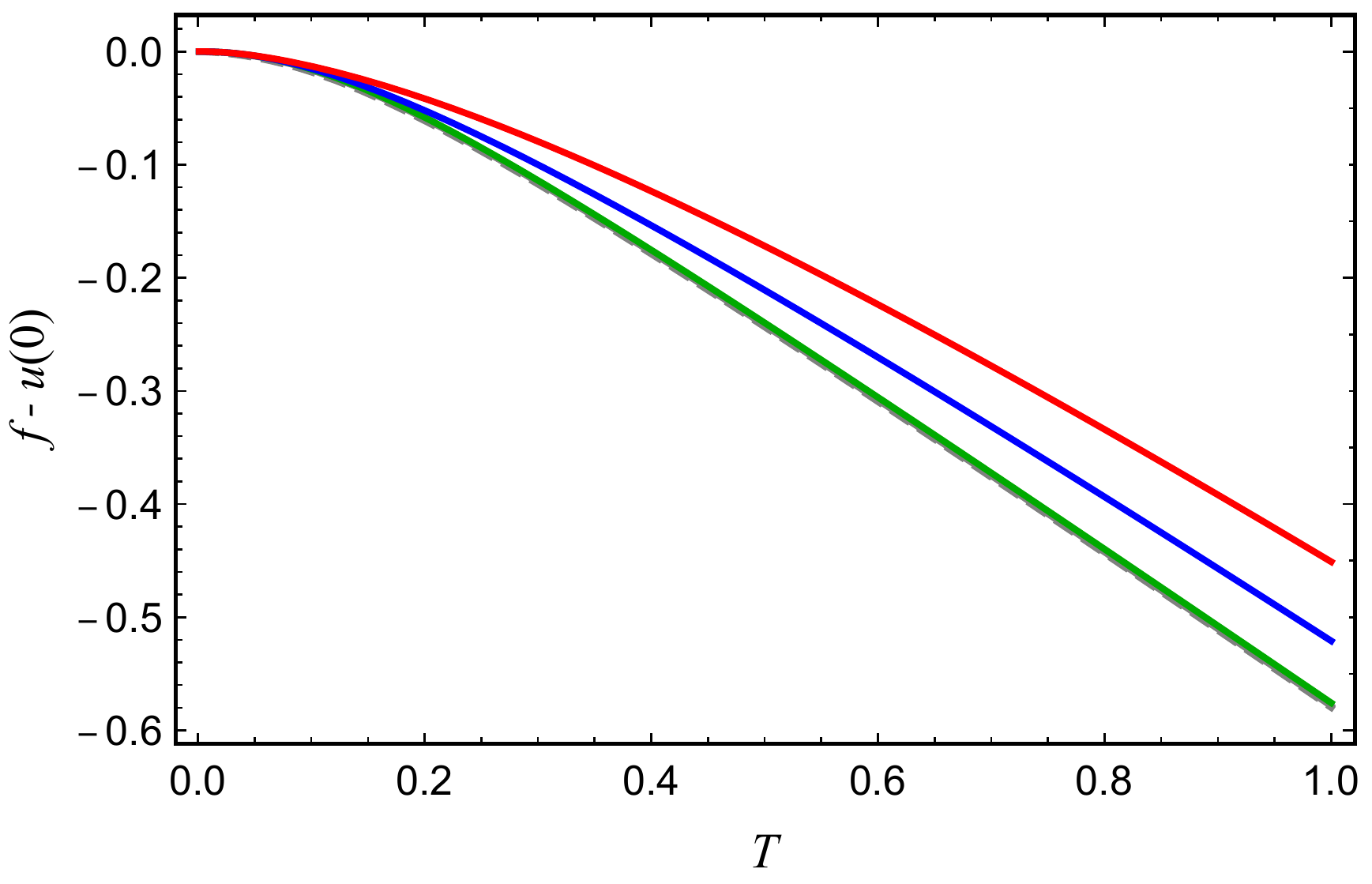}\hfill
  \includegraphics[height=.31\textwidth]{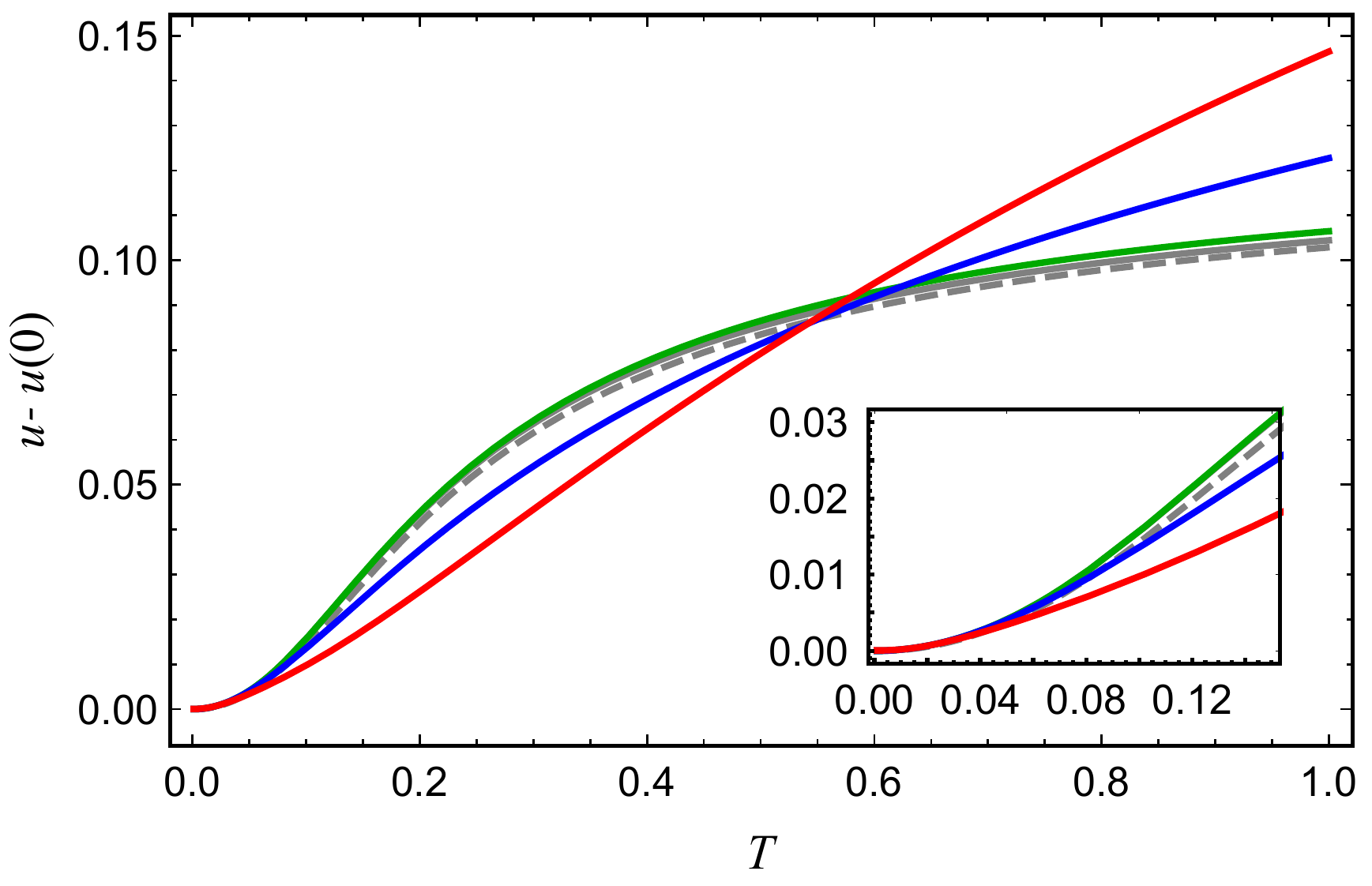}\\[2mm]
  \includegraphics[height=.31\textwidth]{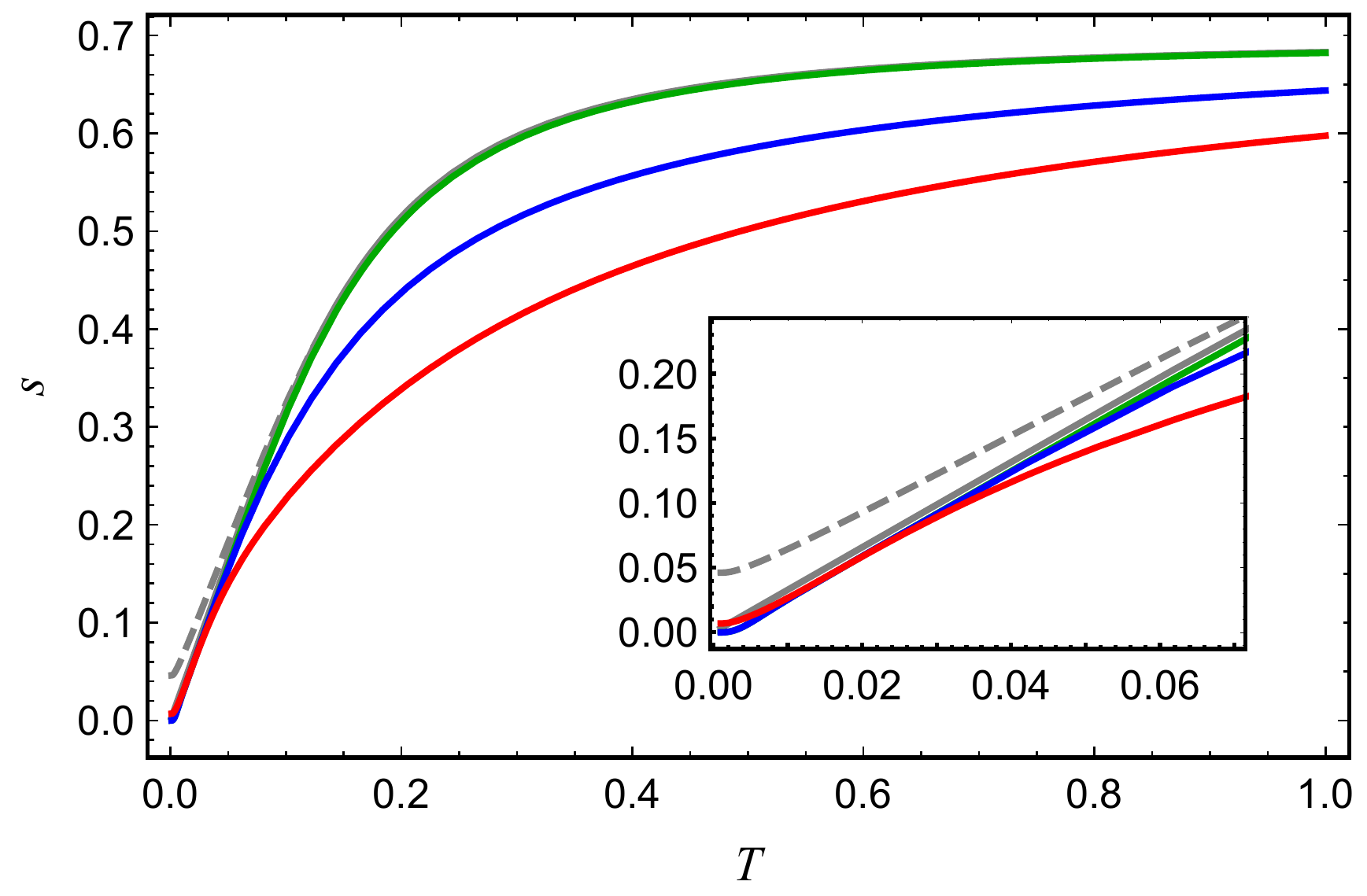}\hfill
  \includegraphics[height=.31\textwidth]{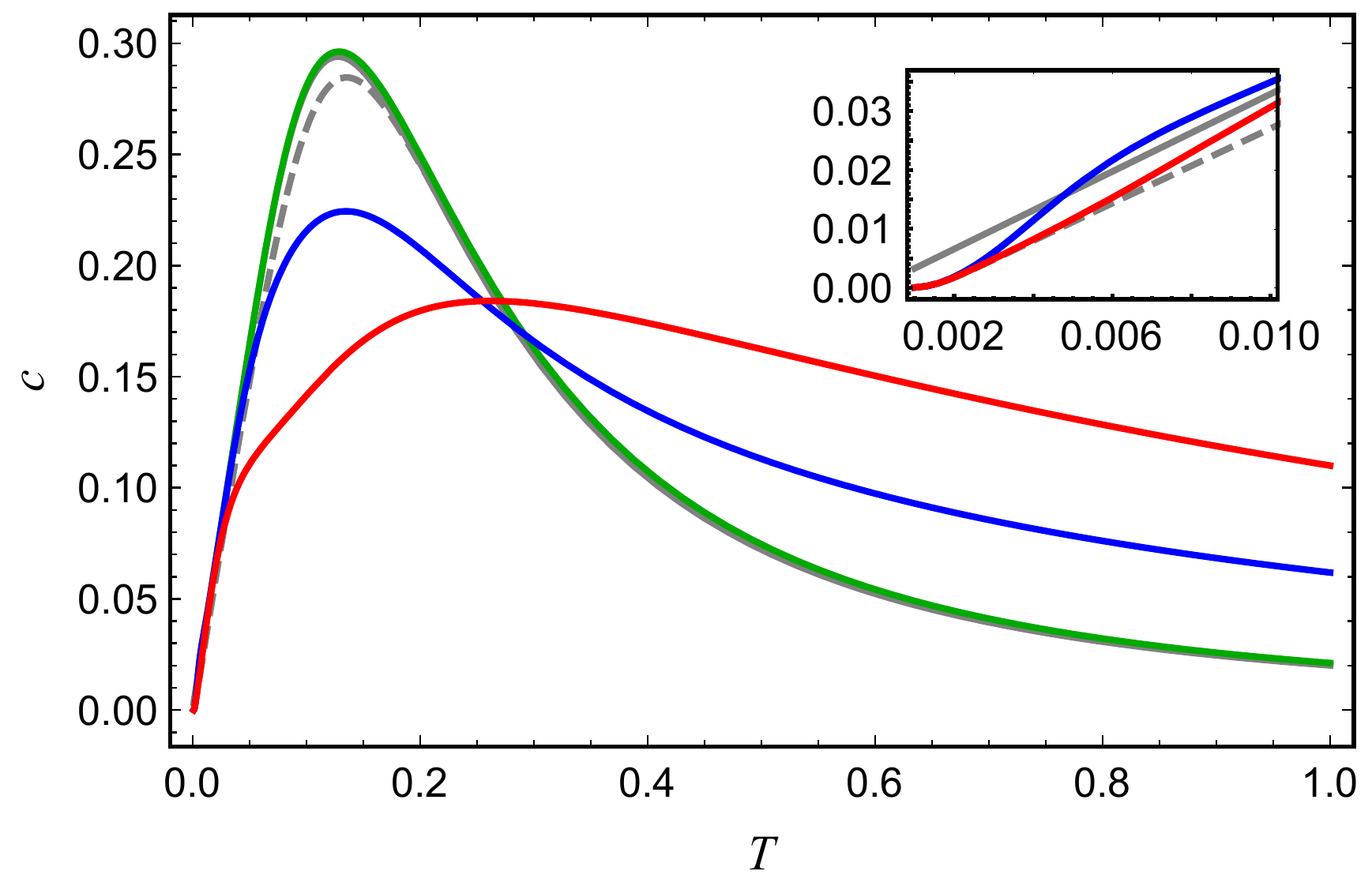}\caption{Thermodynamic functions for the $\su(2)$
    ferromagnetic LMG-PF model with $h(N_1,N_2)=[(N_1-n_1N)^2+(N_2-n_2N)^2]/N$ for $N=100$ spins.
    The red, blue and green lines correspond respectively to the magnon
    densities~$(n_1,n_2)=(1/8,7/8)$, $(1/4,3/4)$ and~$(1/2,1/2)$, while the continuous gray line
    represents the~$h=0$, $N=\infty$ exact result. (In all cases, we have used natural units
    $\hbar=2M=k_{\mathrm B}=1$.)}
  \label{fig.thermo2}
\end{figure}

We shall next analyze the thermodynamics of a class of LMG models of HS type whose deformation
Hamiltonian~\eqref{H1} is given by
\begin{equation}\label{hquad}
h(x_1,\dots,x_m)=\frac1N\,\sum_{a=1}^m(x_a-n_aN)^2\,,
\end{equation}
where the parameters~$n_a$ ($1\le a\le m$) are assumed to lie in the interval~$(0,1)$ and
$n_1+\cdots+n_m=1$. These parameters thus represent the magnon densities of the ground state in
the ferromagnetic case~($\ep=1$). The motivation for considering a quadratic deformation
Hamiltonian is, first of all, that in the original, isotropic LMG model the external term~$H_1$ is
precisely of this form. More recently, generalized LMG models with a quadratic external term have
proved of interest in the context of quantum information theory, since they are some of the few
systems for which the bipartite entanglement entropy of the ground state can be computed in closed
form~\cite{LORV05,ODV08,CFGRT16-LMG}. Using the exact formulas~\eqref{ZTaf}-\eqref{Zferro}
and~\eqref{ZTPF}, we have evaluated the partition function of this class of models for a
relatively large number of spins, of the order of 100 (resp.~$50$) for the $\su(2)$
(resp.~$\su(3)$) ferromagnetic LMG-PF models. From the resulting expression, we have computed the
free energy $f$, the internal energy $u$, the entropy $s$ and the specific heat $c$ (per spin, in
all cases) via the formulas
\begin{eqnarray}\fl
  f(T)&=-\frac TN\,\log Z(T)\,,\qquad &u(T)=\frac{T^2}N\,\frac{\partial\log Z(T)}{\partial T}\,,\\
  \fl s(T)&=\frac{\partial}{\partial T}\bigg(\frac TN\log Z(T)\bigg)\,,\qquad
            &c(T)=\frac{2T}N\,\frac{\partial\log Z(T)}{\partial T}+\frac{T^2}N
                  \frac{\partial^2\log Z(T)}{\partial T^2}\,,
\end{eqnarray}
where we have taken Boltzmann's constant $k_{\mathrm B}=1$. We have first verified that the
thermodynamic functions are practically independent of $N$ for $N\lesssim 100$ (in the $\su(2)$
case) and $N\lesssim 50$ (in the $\su(3)$ case). Thus the thermodynamic functions for $N=100$ (in
the $\su(2)$ case) and $N=50$ (in the $\su(3)$ case) can be regarded as a reasonable approximation
of their $N\to\infty$ counterparts. As an additional check, we have compared the results for the
$\su(2)$ PF chain with no deformation Hamiltonian and $N=100$ spins with the exact $N\to\infty$
formulas derived in Ref.~\cite{EFG12}, finding them in excellent agreement
(cf.~Fig.~\ref{fig.thermo2}). In particular, the extensive behavior of the \emph{thermodynamic}
entropy contrasts with the logarithmic growth of the ground-state \emph{entanglement} entropy of
the ferromagnetic ``quadratic'' gLMG models studied in Ref.~\cite{CFGRT16-LMG}.
\begin{figure}[h]
    \includegraphics[height=.31\textwidth]{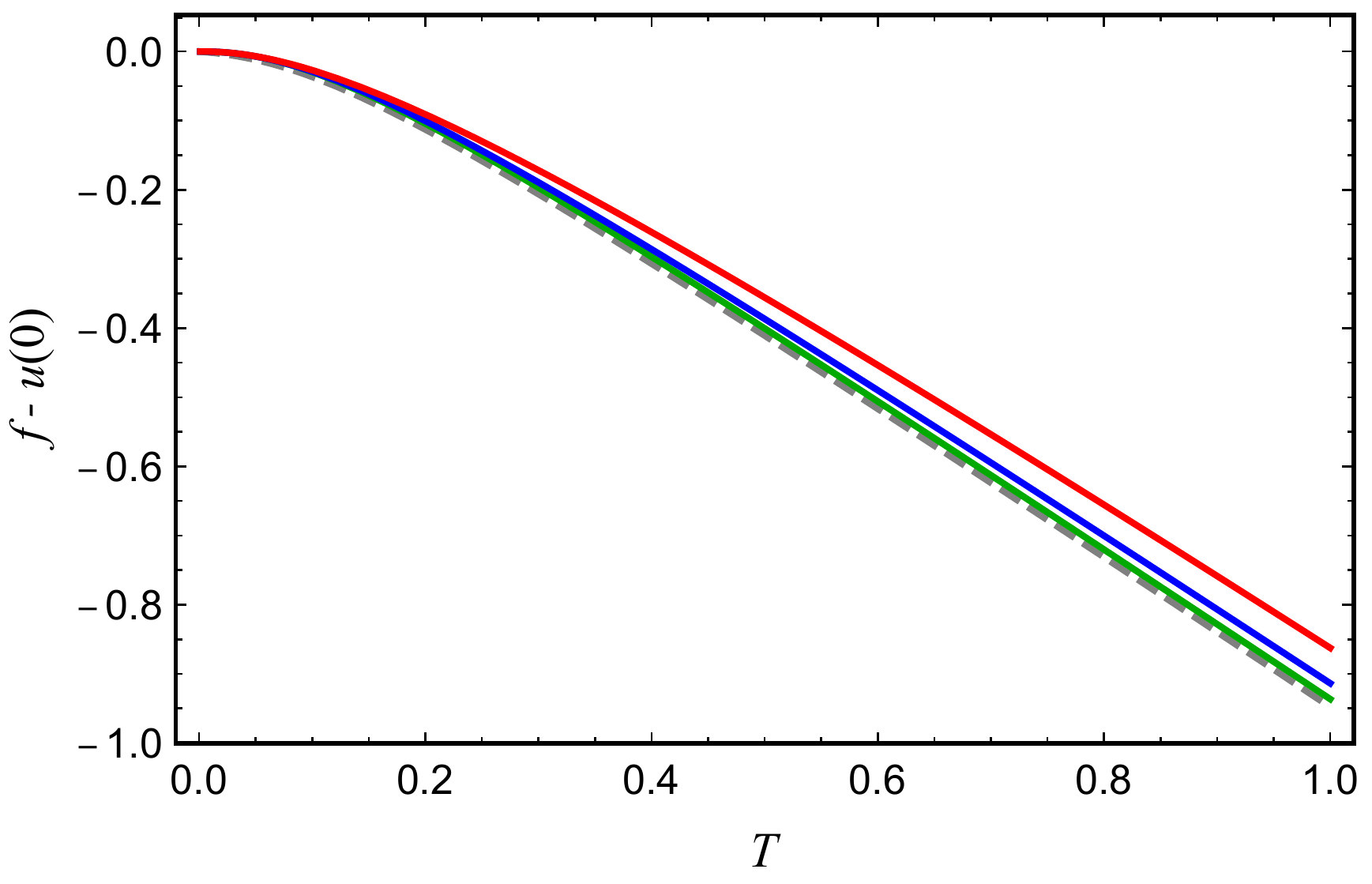}\hfill
    \includegraphics[height=.31\textwidth]{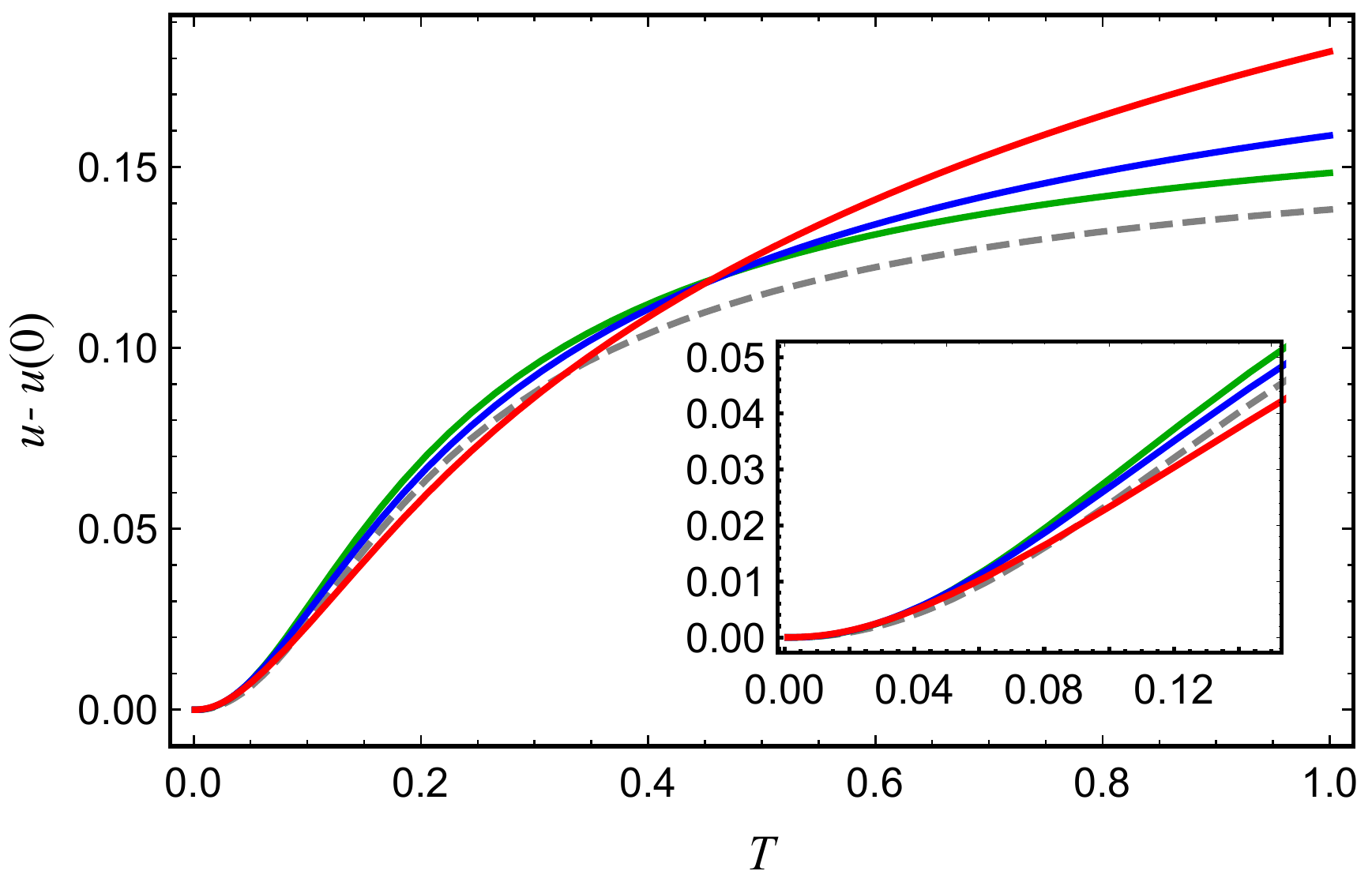}\\[2mm]
    \includegraphics[height=.31\textwidth]{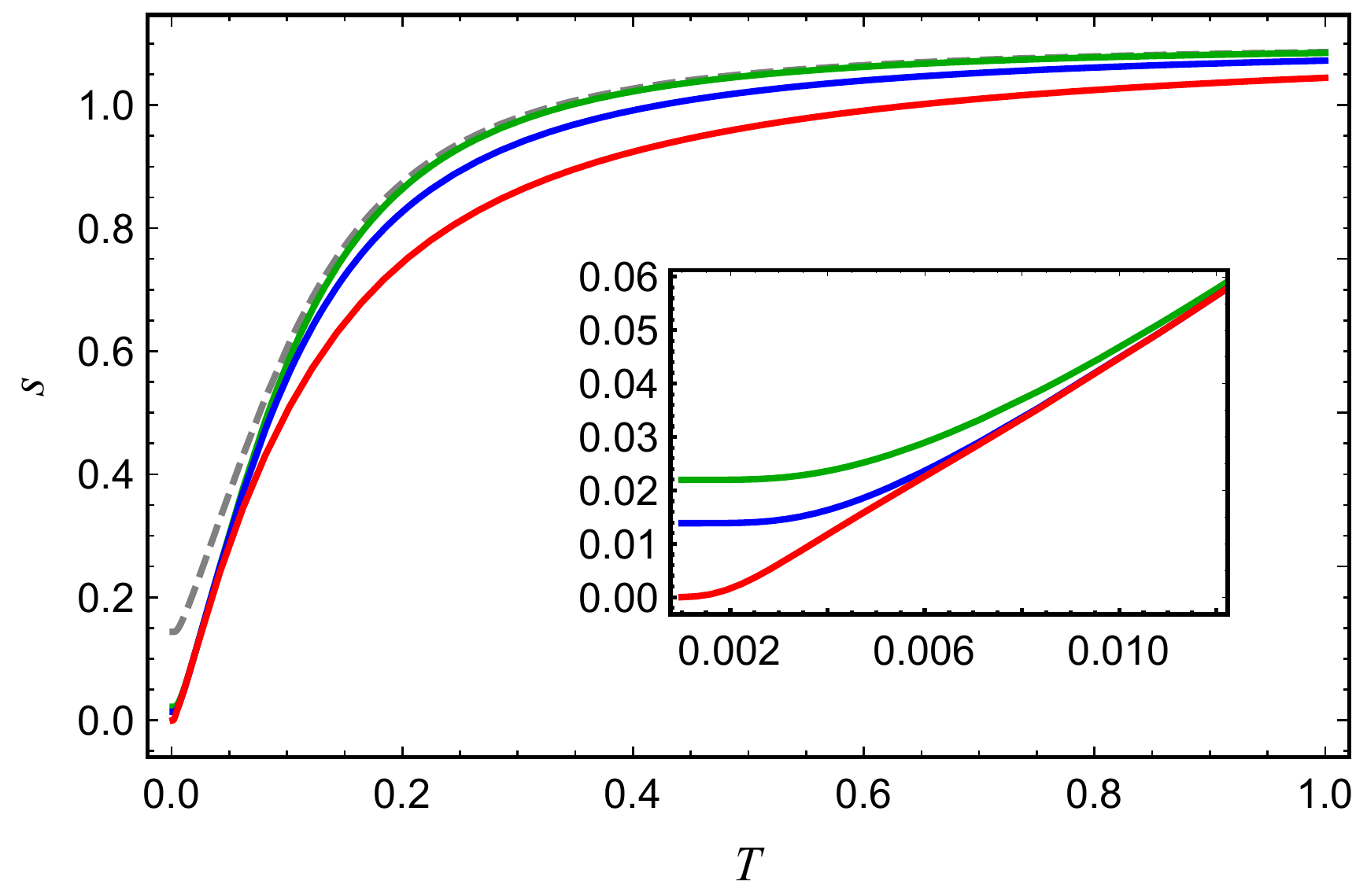}\hfill
    \includegraphics[height=.31\textwidth]{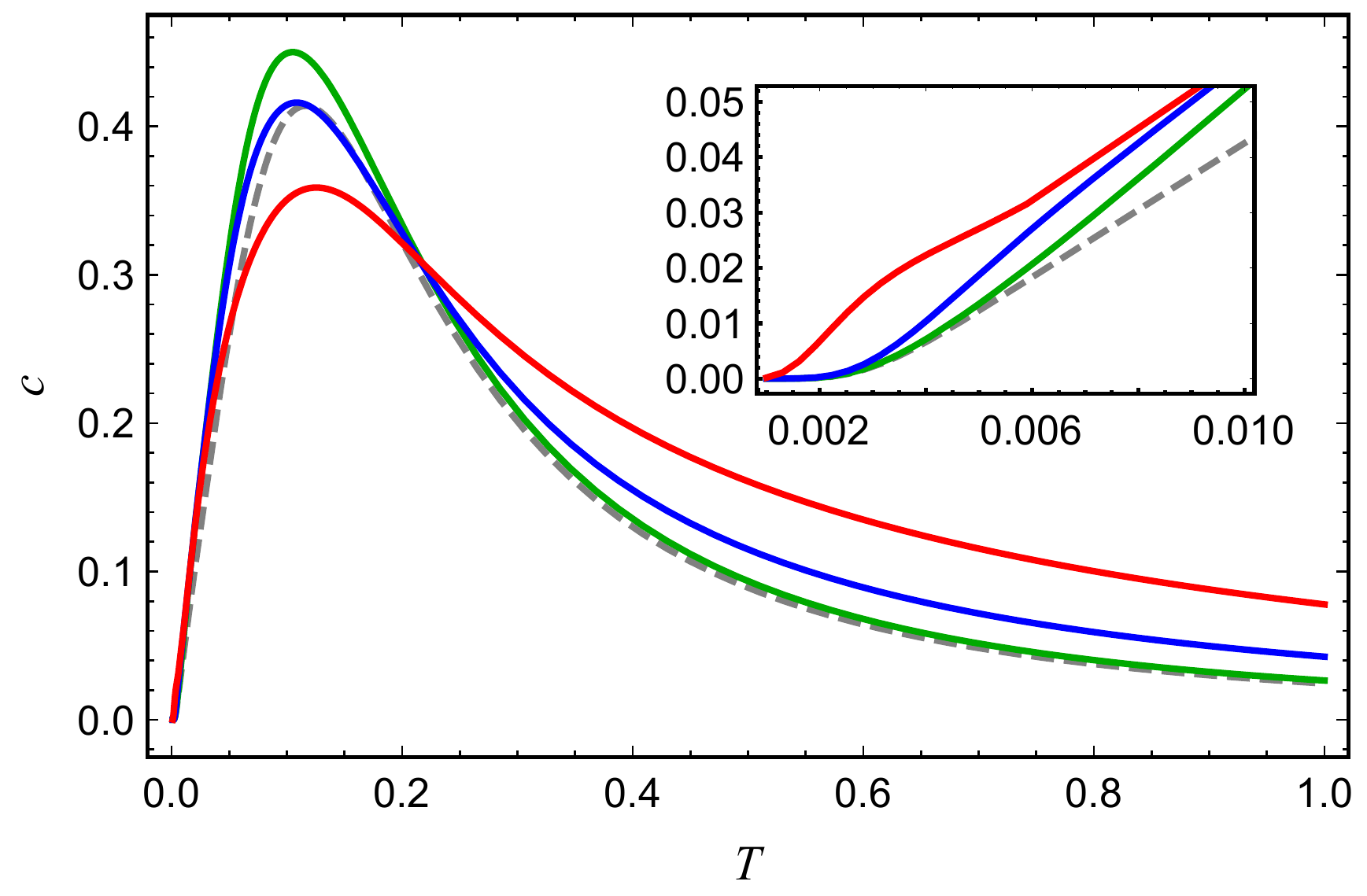}
    \caption{Thermodynamic functions for the $\su(3)$ ferromagnetic LMG-PF model with
      $h(N_1,N_2,N_3)=\sum_{i=1}^3(N_i-n_iN)^2/N$ for $N=50$ spins. The red, blue and green lines
      correspond respectively to the magnon densities~$(n_1,n_2,n_3)=(1/8,1/4,5/8)$,
      $(1/4,1/4,1/2)$ and~$(1/3,1/3,1/3)$. (In all cases, we have used natural units
    $\hbar=2M=k_{\mathrm B}=1$.)}
  \label{fig.thermo3}
\end{figure}

In Figs.~\ref{fig.thermo2} and~\ref{fig.thermo3} we present the plots of the free and internal
energies, the entropy and the specific heat (per spin) respectively of the $\su(2)$ and~$\su(3)$
models~\eqref{H}-\eqref{hquad} in the PF case. It is
apparent from these figures that both the~$\su(2)$ and the $\su(3)$ thermodynamic functions
qualitatively behave like those of a two-level system, as for instance the one-dimensional Ising
model at zero magnetic field or a paramagnetic spin $1/2$ ion~\cite{Mu10}. In particular, from
Figs.~\ref{fig.thermo2} and~\ref{fig.thermo3} we see that the specific heat exhibits the Schottky
peak characteristic of the latter systems. Finally, it may seem surprising that the entropy per
spin does not appear to vanish at~$T=0$ in some cases, especially when $h=0$ (see, e.g.,
Fig.~\ref{fig.thermo3}). Of course, the explanation for this behavior is that the number of spins
$N$ is finite (though large), so that $s(0)=(\log d(m,N))/N$, where $d(m,N)$ is the ground state
degeneracy. In the ferromagnetic case under consideration, it follows from Eq.~\eqref{H} that
when~$h=0$ the ground states are the symmetric states, so that
\[
  d(m,N)=\binom{N+m-1}{m-1}\simeq\frac{N^{m-1}}{(m-1)!}\,,
\]
and thus~$s(0)\simeq(m-1)(\log N)/N$ is small but nonzero. On the other hand, when $h$ does not
vanish identically the $H_1$~term in Eq.~\eqref{H} breaks the ground state degeneracy almost
completely (the more so in the less symmetric cases, in which the densities~$n_a$ are all
different), so that~$s(0)$ is significantly smaller than its $h=0$ counterpart.

\section{Conclusions}

We shall finish this paper with a brief summary of its main results. We have introduced a family
of generalized $\su(m)$ Lipkin--Meshkov--Glick models whose interacting term is a spin chain of
Haldane--Shastry type, which can be equivalently regarded as the deformation of a spin chain of HS
type $H_0$ by the addition of a term $H_1$ in the enveloping algebra of the Cartan subalgebra of
$\su(m)$. The Hilbert space of the system is a direct sum of subspaces $\cH(\bN)$ with fixed
magnon numbers, in which the action of the deformation term is diagonal, so that the model's
partition function decomposes as in Eq.~\eqref{ZTfinal}. By a suitable adaptation of
Polychronakos's freezing trick, we have been able to compute in closed form the partition
functions of the restrictions of the spin chain Hamiltonian $H_0$ to the subspaces $\cH(\bN)$. In
view of the previous remarks, this immediately yields the partition function of the associated
gLMG model. In particular, when~$H_0$ is the Hamiltonian of the Polychronakos--Frahm spin chain we
have obtained an alternative, simpler expression for the partition function akin to
Polychronakos's formula~\cite{Po94} for the case~$H_1=0$. This closed-form expression for the
partition function of the restriction of~$H_0$ to the subspaces $\cH(\bN)$ has been used in
numerical calculations to provide strong evidence that the level density of the latter restriction
is Gaussian when the number of spins tends to infinity. In view of the results of
Ref.~\cite{EFG10}, this suggests that there exists a description of the spectrum of
$H_0\big|_{\cH(\bN)}$ in terms of motifs, a fact that deserves further investigation. We have also
numerically studied the distribution of the spacings of consecutive unfolded levels
of~$H_0\big|_{\cH(\bN)}$, showing that it follows the same characteristic law previously found for
the complete spectrum. As a final application, we have computed the free and internal energies,
the entropy and the specific heat per spin of a class of~$\su(2)$ and~$\su(3)$ gLMG models with
quadratic~$H_1$. We have checked that these functions are virtually independent of the number of
spins~$N$ when this number is sufficiently large, which indicates that they yield reasonable
approximations to their respective thermodynamic limits. Our analysis shows that the thermodynamic
functions of these models are qualitatively similar to those of a two-level system, as already
observed in Ref.~\cite{EFG12} for the~$\su(2)$ chains of HS type. In the latter chains, this
similarity is ultimately due to the existence of a description of the spectrum in terms of motifs,
which leads to simple closed formulas for the thermodynamic functions in terms of the dispersion
relation. This again suggests that a description of this type should also exist for the more
general models studied in this paper.

\section*{References}


\begin{thebibliography}{10}
\providecommand{\url}[1]{\texttt{#1}}
\providecommand{\urlprefix}{URL }
\providecommand{\eprint}[2][]{\url{#2}}

\bibitem{LMG65}
Lipkin H~J, Meshkov N and Glick A~J, \emph{Validity of many-body approximation
  methods for a solvable model: ({I}). {E}xact solutions and perturbation
  theory,}  1965 \emph{Nucl. Phys.} \textbf{62} 188

\bibitem{MGL65}
Meshkov N, Glick A~J and Lipkin H~J, \emph{Validity of many-body approximation
  methods for a solvable model: ({II}). {L}inearization procedures,}  1965
  \emph{Nucl. Phys.} \textbf{62} 199

\bibitem{GLM65}
Glick A~J, Lipkin H~J and Meshkov N, \emph{Validity of many-body approximation
  methods for a solvable model: ({III}). {D}iagram summations,}  1965
  \emph{Nucl. Phys.} \textbf{62} 211

\bibitem{RS80}
Ring P and Schuck P, \emph{The Nuclear Many-Body Problem} (Berlin:
  Springer-Verlag), first edition 1980

\bibitem{UF03}
Unanyan R~G and Fleischhauer M, \emph{{D}ecoherence-free generation of
  many-particle entanglement by adiabatic ground-state transitions,}  2003
  \emph{Phys. Rev. Lett.} \textbf{90} 133601(4)

\bibitem{CLJ09}
Chen G, Liang J~Q and Jia S, \emph{{I}nteraction-induced
  {L}ipkin--{M}eshkov--{G}lick model in a {B}ose--{E}instein condensate inside
  an optical cavity,}  2009 \emph{Opt. Express} \textbf{17} 19682

\bibitem{UZ92}
Ulyanov V~V and Zaslavskii O~B, \emph{{N}ew methods in the theory of quantum
  spin chains,}  1992 \emph{Phys. Rep.} \textbf{216} 179

\bibitem{ORTR16}
Opanchuk B, Rosales-Zárate L, Teh R~Y and Reid M~D, \emph{Quantifying the
  mesoscopic quantum coherence of approximate {NOON} states and spin-squeezed
  two-mode {B}ose--{E}instein condensates,}  2016 \emph{Phys. Rev. A}
  \textbf{94} 062125(14)

\bibitem{RCCP17}
Romera E, Castaños O, Calixto M and Pérez-Bernal F, \emph{Delocalization
  properties at isolated avoided crossings in {L}ipkin--{M}eshkov--{G}lick type
  {H}amiltonian models,}  2017 \emph{J. Stat. Mech. Theory-E.} \textbf{2017}
  P013101(19)

\bibitem{PS05}
Popkov V and Salerno M, \emph{Logarithmic divergence of the block entanglement
  entropy for the ferromagnetic {H}eisenberg model,}  2005 \emph{Phys. Rev. A}
  \textbf{71} 012301(4)

\bibitem{LORV05}
Latorre J~I, Or{\'u}s R, Rico E and Vidal J, \emph{Entanglement entropy in the
  {L}ipkin--{M}eshkov--{G}lick model,}  2005 \emph{Phys. Rev. A} \textbf{71}
  064101(4)

\bibitem{BDV06}
Barthel T, Dusuel S and Vidal J, \emph{Entanglement entropy beyond the free
  case,}  2006 \emph{Phys. Rev. Lett.} \textbf{97} 220402(4)

\bibitem{ODV08}
Orus R, Dusuel S and Vidal J, \emph{Equivalence of critical scaling laws for
  many-body entanglement in the {L}ipkin--{M}eshkov--{G}lick model,}  2008
  \emph{Phys. Rev. Lett.} \textbf{101} 25701(4)

\bibitem{WVVD12}
Wilms J, Vidal J, Verstraete F and Dusuel S, \emph{Finite-temperature mutual
  information in a simple phase transition,}  2012 \emph{J. Stat. Mech.-Theory
  E.} \textbf{2012} P01023(21)

\bibitem{VLRK03}
Vidal G, Latorre J~I, Rico E and Kitaev A, \emph{Entanglement in quantum
  critical phenomena,}  2003 \emph{Phys. Rev. Lett.} \textbf{90} 227902(4)

\bibitem{HLW94}
Holzhey C, Larsen F and Wilczek F, \emph{Geometric and renormalized entropy in
  conformal field theory,}  1994 \emph{Nucl. Phys. B} \textbf{424} 443

\bibitem{Ko04}
Korepin V~E, \emph{Universality of entropy scaling in one dimensional gapless
  models,}  2004 \emph{Phys. Rev. Lett.} \textbf{92} 096402(3)

\bibitem{RM04}
Refael G and Moore J~E, \emph{Entanglement entropy of random quantum critical
  points in one dimension,}  2004 \emph{Phys. Rev. Lett.} \textbf{93} 260602(4)

\bibitem{CFGRT16-LMG}
Carrasco J~A, Finkel F, Gonz{\'a}lez-L{\'o}pez A, Rodr{\'\i}guez M~A and
  Tempesta P, \emph{Generalized isotropic {L}ipkin--{M}eshkov--{G}lick models:
  ground state entanglement and quantum entropies,}  2016 \emph{J. Stat. Mech.
  Theory-E.} \textbf{2016} 033114(33)

\bibitem{PD99}
Pan F and Draayer J~P, \emph{Analytical solutions for the {LMG} model,}  1999
  \emph{Phys. Lett. B} \textbf{451} 1

\bibitem{MOPN06}
Morita H, Ohnishi H, da~Provid{\^{e}}ncia J and Nishiyama S, \emph{Exact
  solutions for the {LMG} model {H}amiltonian based on the {B}ethe ansatz,}
  2006 \emph{Nucl. Phys. B} \textbf{737} 337

\bibitem{RVM07}
Ribeiro P, Vidal J and Mosseri R, \emph{Thermodynamical limit of the
  {L}ipkin--{M}eshkov--{G}lick model,}  2007 \emph{Phys. Rev. Lett.}
  \textbf{99} 050402(4)

\bibitem{RVM08}
Ribeiro P, Vidal J and Mosseri R, \emph{Exact spectrum of the
  {L}ipkin--{M}eshkov--{G}lick model in the thermodynamic limit and finite-size
  corrections,}  2008 \emph{Phys. Rev. E} \textbf{78} 021106(13)

\bibitem{Ts88}
Tsallis C, \emph{Possible generalization of {B}oltzmann--{G}ibbs statistics,}
  1988 \emph{J. Stat. Phys.} \textbf{52} 479

\bibitem{Ts09}
Tsallis C, \emph{Introduction to Nonextensive Statistical Mechanics:
  Approaching a Complex World} (Berlin: Springer) 2009

\bibitem{Ha88}
Haldane F~D~M, \emph{{E}xact {J}astrow--{G}utzwiller resonating-valence-bond
  ground state of the spin-$1/2$ antiferromagnetic {H}eisenberg chain with
  $1/r^2$ exchange,}  1988 \emph{Phys. Rev. Lett.} \textbf{60} 635

\bibitem{Sh88}
Shastry B~S, \emph{{E}xact solution of an ${S}=1/2$ {H}eisenberg
  antiferromagnetic chain with long-ranged interactions,}  1988 \emph{Phys.
  Rev. Lett.} \textbf{60} 639

\bibitem{HHTBP92}
Haldane F~D~M, Ha Z~N~C, Talstra J~C, Bernard D and Pasquier V, \emph{{Y}angian
  symmetry of integrable quantum chains with long-range interactions and a new
  description of states in conformal field theory,}  1992 \emph{Phys. Rev.
  Lett.} \textbf{69} 2021

\bibitem{Po93}
Polychronakos A~P, \emph{{L}attice integrable systems of {H}aldane--{S}hastry
  type,}  1993 \emph{Phys. Rev. Lett.} \textbf{70} 2329

\bibitem{Fr93}
Frahm H, \emph{{S}pectrum of a spin chain with inverse-square exchange,}  1993
  \emph{J. Phys. A: Math. Gen.} \textbf{26} L473

\bibitem{FI94}
Frahm H and Inozemtsev V~I, \emph{New family of solvable 1{D} {H}eisenberg
  models,}  1994 \emph{J. Phys. A: Math. Gen.} \textbf{27} L801

\bibitem{Gu63}
Gutzwiller M~C, \emph{{E}ffect of correlation on the ferromagnetism of
  transition metals,}  1963 \emph{Phys. Rev. Lett.} \textbf{10} 159

\bibitem{GV87}
Gebhard F and Vollhardt D, \emph{{C}orrelation functions for {H}ubbard-type
  models: the exact results for the {G}utzwiller wave function in one
  dimension,}  1987 \emph{Phys. Rev. Lett.} \textbf{59} 1472

\bibitem{Ha91b}
Haldane F~D~M, \emph{``{F}ractional statistics'' in arbitrary dimensions: a
  generalization of the {P}auli principle,}  1991 \emph{Phys. Rev. Lett.}
  \textbf{67} 937

\bibitem{GS05}
Greiter M and Schuricht D, \emph{No attraction between spinons in the
  {H}aldane--{S}hastry model,}  2005 \emph{Phys. Rev. B} \textbf{71} 224424(4)

\bibitem{Gr09}
Greiter M, \emph{{S}tatistical phases and momentum spacings for one-dimensional
  anyons,}  2009 \emph{Phys. Rev. B} \textbf{79} 064409(5)

\bibitem{BBS08}
Basu-Mallick B, Bondyopadhaya N and Sen D, \emph{{L}ow energy properties of the
  {$\mathrm{SU}(m|n)$} supersymmetric {H}aldane--{S}hastry spin chain,}  2008
  \emph{Nucl. Phys. B} \textbf{795} 596

\bibitem{CS10}
Cirac J~I and Sierra G, \emph{Infinite matrix product states, conformal field
  theory, and the {H}aldane--{S}hastry model,}  2010 \emph{Phys. Rev. B}
  \textbf{81} 104431(4)

\bibitem{BBL08}
Bargheer T, Beisert N and Loebbert F, \emph{Boosting nearest-neighbour to
  long-range integrable spin chains,}  2008 \emph{J. Stat. Mech. Theory-E.}
  \textbf{2008} L11001(9)

\bibitem{BBL09}
Bargheer T, Beisert N and Loebbert F, \emph{Long-range deformations for
  integrable spin chains,}  2009 \emph{J. Phys. A: Math. Theor.} \textbf{42}
  285205(58)

\bibitem{HGCK16}
Hung C~L, Gonz{\'a}lez-Tudela A, Cirac J~I and Kimble H~J, \emph{Quantum spin
  dynamics with pairwise-tunable, long-range interactions,}  2016 \emph{Proc.
  Natl. Acad. Sci. U. S. A.} \textbf{113} E4946

\bibitem{Po94}
Polychronakos A~P, \emph{{E}xact spectrum of {$\mathrm{SU}(n)$} spin chain with
  inverse-square exchange,}  1994 \emph{Nucl. Phys. B} \textbf{419} 553

\bibitem{FG05}
Finkel F and Gonz{\'a}lez-L{\'o}pez A, \emph{{G}lobal properties of the
  spectrum of the {H}aldane--{S}hastry spin chain,}  2005 \emph{Phys. Rev. B}
  \textbf{72} 174411(6)

\bibitem{BFGR10}
Barba J~C, Finkel F, Gonz\'alez-L\'opez A and Rodr{\'\i}guez M~A,
  \emph{Inozemtsev's hyperbolic spin model and its related spin chain,}  2010
  \emph{Nucl. Phys. B} \textbf{839} 499

\bibitem{Ca71}
Calogero F, \emph{{S}olution of the one-dimensional ${N}$-body problems with
  quadratic and/or inversely quadratic pair potentials,}  1971 \emph{J. Math.
  Phys.} \textbf{12} 419

\bibitem{Su71}
Sutherland B, \emph{{E}xact results for a quantum many-body problem in one
  dimension,}  1971 \emph{Phys. Rev. A} \textbf{4} 2019

\bibitem{Su72}
Sutherland B, \emph{{E}xact results for a quantum many-body problem in one
  dimension. {I}{I},}  1972 \emph{Phys. Rev. A} \textbf{5} 1372

\bibitem{In96}
Inozemtsev V~I, \emph{Exactly solvable model of interacting electrons confined
  by the {M}orse potential,}  1996 \emph{Phys. Scr.} \textbf{53} 516

\bibitem{BT77}
Berry M~V and Tabor M, \emph{{L}evel clustering in the regular spectrum,}  1977
  \emph{Proc. R. Soc. London Ser. A} \textbf{356} 375

\bibitem{BGS84}
Bohigas O, Giannoni M~J and Schmit C, \emph{{C}haracterization of chaotic
  quantum spectra and universality of level fluctuation laws,}  1984
  \emph{Phys. Rev. Lett.} \textbf{52} 1

\bibitem{BFGR08epl}
Barba J~C, Finkel F, Gonz\'alez-L\'opez A and Rodr{\'\i}guez M~A, \emph{{T}he
  {B}erry--{T}abor conjecture for spin chains of {H}aldane--{S}hastry type,}
  2008 \emph{Europhys. Lett.} \textbf{83} 27005(6)

\bibitem{EFG12}
Enciso A, Finkel F and Gonz{\'a}lez-L{\'o}pez A, \emph{Thermodynamics of spin
  chains of {H}aldane--{S}hastry type and one-dimensional vertex models,}  2012
  \emph{Ann. Phys.-New York} \textbf{327} 2627

\bibitem{EFG09}
Enciso A, Finkel F and Gonz{\'a}lez-L{\'o}pez A, \emph{Spin chains of
  {H}aldane--{S}hastry type and a generalized central limit theorem,}  2009
  \emph{Phys. Rev. E} \textbf{79} 060105(4)

\bibitem{EFG10}
Enciso A, Finkel F and Gonz{\'a}lez-L{\'o}pez A, \emph{Level density of spin
  chains of {H}aldane--{S}hastry type,}  2010 \emph{Phys. Rev. E} \textbf{82}
  051117(6)

\bibitem{BBH10}
Basu-Mallick B, Bondyopadhaya N and Hikami K, \emph{One-dimensional vertex
  models associated with a class of {Y}angian invariant {H}aldane--{S}hastry
  like spin chains,}  2010 \emph{SIGMA} \textbf{6} 091(13)

\bibitem{Ha01}
Haake F, \emph{{Q}uantum {S}ignatures of {C}haos} (Berlin: Springer-Verlag),
  second edition 2001

\bibitem{BFGR08}
Barba J~C, Finkel F, Gonz\'alez-L\'opez A and Rodr{\'\i}guez M~A,
  \emph{{P}olychronakos--{F}rahm spin chain of {$BC_N$} type and the
  {B}erry--{T}abor conjecture,}  2008 \emph{Phys. Rev. B} \textbf{77}
  214422(10)

\bibitem{BFGR09}
Barba J~C, Finkel F, Gonz\'alez-L\'opez A and Rodr{\'\i}guez M~A, \emph{{A}n
  exactly solvable supersymmetric spin chain of {$BC_N$} type,}  2009
  \emph{Nucl. Phys. B} \textbf{806} 684

\bibitem{Mu10}
Mussardo G, \emph{Statistical Field Theory: an Introduction to Exactly Solved
  Models in Statistical Physics} (Oxford: Oxford University Press) 2010

\end{thebibliography}

\end{document}